\documentclass[11pt]{article}
\usepackage{graphicx}
\usepackage{amssymb}
\usepackage{amsmath}
\usepackage{booktabs}
\usepackage{multirow}
\usepackage[title]{appendix}
\usepackage[title]{appendix}
\usepackage{multirow}
\usepackage{floatrow}
\usepackage{hyperref}
\usepackage{xcolor}
\usepackage{subcaption}
\usepackage{cite}
\usepackage{xspace}
\usepackage[normalem]{ulem}

\newcommand{\pp}      {$K^+\rightarrow\pi^+\pi^0$}

\newcommand{\pgg}     {$K^+\rightarrow\pi^+\gamma\gamma$~}

\newcommand{\pppc}    {$K^+\rightarrow\pi^+\pi^+\pi^-$}
\newcommand{\klthree} {$K^+\rightarrow\pi^0l^+\nu$~}
\newcommand{\kethree} {$K^+\rightarrow\pi^0e^+\nu$}
\newcommand{\kefour}  {$K^+\rightarrow \pi^+\pi^- e^+\nu$}
\newcommand{\mn}      {$K^+\rightarrow\mu^+\nu$}

\newcommand{\ppgr}    {$K^+\rightarrow\pi^+\pi^0\gamma$~}
\newcommand{\pnnc}    {$K^+\rightarrow\pi^+\nu\bar{\nu}$~}
\newcommand{\pnnz}    {$K_L\rightarrow\pi^0\nu\bar{\nu}$~}

\def\Journal#1#2#3#4{{#1} {\bf #2}, #3 (#4)}

\def\NIMA{{\em Nucl. Instrum. Methods} A}
\def\JINST{{\em Journal of Instrum.}}
\def\JINST{{\em JINST}}
\def\JHEP{{\em JHEP}}
\def\EPJ{{\em Eur. Phys. J.} C}
\def\IJMP{{\em Int. J. Mod. Phys.} A}
\def\NPB{{\em Nucl. Phys.} B}
\def\PLB{{\em Phys. Lett.}  B}
\def\PRL{\em Phys. Rev. Lett.}
\def\PRD{{\em Phys. Rev.} D}

\def\JPG{{\em J. Phys.} G}
\def\PTEP{{\em Prog. Theor. Exp. Phys.}}

\newfloatcommand{capbtabbox}{table}[][\FBwidth]

\begin{document}

\pagenumbering{arabic}
\centerline{\LARGE EUROPEAN ORGANIZATION FOR NUCLEAR RESEARCH}
\begin{flushright}
CERN-EP-2021-050\\
March 27, 2021 \\
\end{flushright}
\vspace{15mm}

\begin{center}
\Large{\bf Measurement of the very rare \boldmath\pnnc decay \\
\vspace{5mm}
}
The NA62 Collaboration
\end{center}
\vspace{10mm}
\begin{abstract}
  The NA62 experiment reports the branching ratio measurement
  BR$(K^+ \rightarrow \pi^+ \nu\bar{\nu}) = (10.6^{+4.0}_{-3.4} |_{\rm
    stat} \pm 0.9_{\rm syst}) \times 10 ^{-11}$ at 68\% CL, based on
  the observation of 20 signal candidates with an expected background
  of 7.0 events from the total data sample collected at the CERN SPS
  during 2016--2018. This provides evidence for the very rare \pnnc
  decay, observed with a significance of 3.4$\sigma$. The experiment
  achieves a single event sensitivity of
  $(0.839\pm 0.054)\times10^{-11}$, corresponding to 10.0 events
  assuming the Standard Model branching ratio of
  $(8.4\pm1.0)\times10^{-11}$. This measurement is also used to set
  limits on BR($K^+ \to \pi^+ X$), where $X$ is a scalar or
  pseudo-scalar particle. Details are given of the analysis of the
  2018 data sample, which corresponds to about 80\% of the total data
  sample.

\end{abstract}
\vspace{20mm}
\begin{center}
{\it Accepted for publication in  JHEP}
\end{center}
\clearpage

\begin{center}
{\Large The NA62 Collaboration$\,$\renewcommand{\thefootnote}{\fnsymbol{footnote}}%
\footnotemark[1]\renewcommand{\thefootnote}{\arabic{footnote}}}\\
\end{center}
\vspace{3mm}
\begin{raggedright}
\noindent

{\bf Universit\'e Catholique de Louvain, Louvain-La-Neuve, Belgium}\\
 E.~Cortina Gil,
 A.~Kleimenova,
 E.~Minucci$\,$\footnotemark[1]$^,\,$\footnotemark[2],
 S.~Padolski$\,$\footnotemark[3],
 P.~Petrov,
 A.~Shaikhiev$\,$\footnotemark[4],
 R.~Volpe$\,$\footnotemark[5]\\[2mm]

{\bf TRIUMF, Vancouver, British Columbia, Canada}\\
 T.~Numao,
 Y.~Petrov,
 B.~Velghe\\[2mm]

{\bf University of British Columbia, Vancouver, British Columbia, Canada}\\
 D.~Bryman$\,$\footnotemark[6],
 J.~Fu\\[2mm]

{\bf Charles University, Prague, Czech Republic}\\
 T.~Husek$\,$\footnotemark[7],
 J.~Jerhot$\,$\footnotemark[8],
 K.~Kampf,
 M.~Zamkovsky\\[2mm]

{\bf Institut f\"ur Physik and PRISMA Cluster of Excellence, Universit\"at Mainz, Mainz, Germany}\\
 R.~Aliberti$\,$\footnotemark[9],
 G.~Khoriauli$\,$\footnotemark[10],
 J.~Kunze,
 D.~Lomidze$\,$\footnotemark[11],
 L.~Peruzzo,
 M.~Vormstein,
 R.~Wanke\\[2mm]

{\bf Dipartimento di Fisica e Scienze della Terra dell'Universit\`a e INFN, Sezione di Ferrara, Ferrara, Italy}\\
 P.~Dalpiaz,
 M.~Fiorini,
 I.~Neri,
 A.~Norton$\,$\footnotemark[12],
 F.~Petrucci,
 H.~Wahl$\,$\footnotemark[13]\\[2mm]

{\bf INFN, Sezione di Ferrara, Ferrara, Italy}\\
 A.~Cotta Ramusino,
 A.~Gianoli\\[2mm]

{\bf Dipartimento di Fisica e Astronomia dell'Universit\`a e INFN, Sezione di Firenze, Sesto Fiorentino, Italy}\\
 E.~Iacopini,
 G.~Latino,
 M.~Lenti,
 A.~Parenti\\[2mm]

{\bf INFN, Sezione di Firenze, Sesto Fiorentino, Italy}\\
 A.~Bizzeti$\,$\footnotemark[14],
 F.~Bucci\\[2mm]

{\bf Laboratori Nazionali di Frascati, Frascati, Italy}\\
 A.~Antonelli,
 G.~Georgiev$\,$\footnotemark[15],
 V.~Kozhuharov$\,$\footnotemark[15],
 G.~Lanfranchi,
 S.~Martellotti,
 M.~Moulson,
 T.~Spadaro\\[2mm]

{\bf Dipartimento di Fisica ``Ettore Pancini'' e INFN, Sezione di Napoli, Napoli, Italy}\\
 F.~Ambrosino,
 T.~Capussela,
 M.~Corvino$\,$\footnotemark[1],
 D.~Di Filippo,
 P.~Massarotti,
 M.~Mirra,
 M.~Napolitano,
 G.~Saracino\\[2mm]

{\bf Dipartimento di Fisica e Geologia dell'Universit\`a e INFN, Sezione di Perugia, Perugia, Italy}\\
 G.~Anzivino,
 F.~Brizioli$\,$\renewcommand{\thefootnote}{\fnsymbol{footnote}}\footnotemark[1]\renewcommand{\thefootnote}{\arabic{footnote}},
 E.~Imbergamo,
 R.~Lollini,
 R.~Piandani$\,$\footnotemark[16],
 C.~Santoni\\[2mm]

{\bf INFN, Sezione di Perugia, Perugia, Italy}\\
 M.~Barbanera,
 P.~Cenci,
 B.~Checcucci,
 P.~Lubrano,
 M.~Lupi$\,$\footnotemark[17],
 M.~Pepe,
 M.~Piccini\\[2mm]

{\bf Dipartimento di Fisica dell'Universit\`a e INFN, Sezione di Pisa, Pisa, Italy}\\
 F.~Costantini,
 L.~Di Lella$\,$\footnotemark[13],
 N.~Doble$\,$\footnotemark[13],
 M.~Giorgi,
 S.~Giudici,
 G.~Lamanna,
 E.~Lari,
 E.~Pedreschi,
 M.~Sozzi\\[2mm]

{\bf INFN, Sezione di Pisa, Pisa, Italy}\\
 C.~Cerri,
 R.~Fantechi,
 L.~Pontisso,
 F.~Spinella\\[2mm]

{\bf Scuola Normale Superiore e INFN, Sezione di Pisa, Pisa, Italy}\\
 I.~Mannelli\\[2mm]

{\bf Dipartimento di Fisica, Sapienza Universit\`a di Roma e INFN, Sezione di Roma I, Roma, Italy}\\
 G.~D'Agostini,
 M.~Raggi\\[2mm]

{\bf INFN, Sezione di Roma I, Roma, Italy}\\
 A.~Biagioni,
 E.~Leonardi,
 A.~Lonardo,
 P.~Valente,
 P.~Vicini\\[2mm]

{\bf INFN, Sezione di Roma Tor Vergata, Roma, Italy}\\
 R.~Ammendola,
 V.~Bonaiuto$\,$\footnotemark[18],
 A.~Fucci,
 A.~Salamon,
 F.~Sargeni$\,$\footnotemark[19]\\[2mm]

{\bf Dipartimento di Fisica dell'Universit\`a e INFN, Sezione di Torino, Torino, Italy}\\
 R.~Arcidiacono$\,$\footnotemark[20],
 B.~Bloch-Devaux,
 M.~Boretto$\,$\footnotemark[1],
 E.~Menichetti,
 E.~Migliore,
 D.~Soldi\\[2mm]

{\bf INFN, Sezione di Torino, Torino, Italy}\\
 C.~Biino,
 A.~Filippi,
 F.~Marchetto\\[2mm]

{\bf Instituto de F\'isica, Universidad Aut\'onoma de San Luis Potos\'i, San Luis Potos\'i, Mexico}\\
 J.~Engelfried,
 N.~Estrada-Tristan$\,$\footnotemark[21]\\[2mm]

{\bf Horia Hulubei National Institute of Physics for R\&D in Physics and Nuclear Engineering, Bucharest-Magurele, Romania}\\
 A. M.~Bragadireanu,
 S. A.~Ghinescu,
 O. E.~Hutanu\\[2mm]

{\bf Joint Institute for Nuclear Research, Dubna, Russia}\\
 A.~Baeva,
 D.~Baigarashev,
 D.~Emelyanov,
 T.~Enik,
 V.~Falaleev,
 V.~Kekelidze,
 A.~Korotkova,
 L.~Litov$\,$\footnotemark[15],
 D.~Madigozhin,
 M.~Misheva$\,$\footnotemark[22],
 N.~Molokanova,
 S.~Movchan,
 I.~Polenkevich,
 Yu.~Potrebenikov,
 S.~Shkarovskiy,
 A.~Zinchenko$\,$\renewcommand{\thefootnote}{\fnsymbol{footnote}}\footnotemark[2]\renewcommand{\thefootnote}{\arabic{footnote}}\\[2mm]

{\bf Institute for Nuclear Research of the Russian Academy of Sciences, Moscow, Russia}\\
 S.~Fedotov,
 E.~Gushchin,
 A.~Khotyantsev,
 Y.~Kudenko$\,$\footnotemark[23],
 V.~Kurochka,
 M.~Medvedeva,
 A.~Mefodev\\[2mm]

{\bf Institute for High Energy Physics - State Research Center of Russian Federation, Protvino, Russia}\\
 S.~Kholodenko,
 V.~Kurshetsov,
 V.~Obraztsov,
 A.~Ostankov$\,$\renewcommand{\thefootnote}{\fnsymbol{footnote}}\footnotemark[2]\renewcommand{\thefootnote}{\arabic{footnote}},
 V.~Semenov$\,$\renewcommand{\thefootnote}{\fnsymbol{footnote}}\footnotemark[2]\renewcommand{\thefootnote}{\arabic{footnote}},
 V.~Sugonyaev,
 O.~Yushchenko\\[2mm]

{\bf Faculty of Mathematics, Physics and Informatics, Comenius University, Bratislava, Slovakia}\\
 L.~Bician$\,$\footnotemark[1],
 T.~Blazek,
 V.~Cerny,
 Z.~Kucerova\\[2mm]

{\bf CERN,  European Organization for Nuclear Research, Geneva, Switzerland}\\
 J.~Bernhard,
 A.~Ceccucci,
 H.~Danielsson,
 N.~De Simone$\,$\footnotemark[24],
 F.~Duval,
 B.~D\"obrich,
 L.~Federici,
 E.~Gamberini,
 L.~Gatignon$\,$\footnotemark[25],
 R.~Guida,
 F.~Hahn$\,$\renewcommand{\thefootnote}{\fnsymbol{footnote}}\footnotemark[2]\renewcommand{\thefootnote}{\arabic{footnote}},
 E. B.~Holzer,
 B.~Jenninger,
 M.~Koval$\,$\footnotemark[26],
 P.~Laycock$\,$\footnotemark[3],
 G.~Lehmann Miotto,
 P.~Lichard,
 A.~Mapelli,
 R.~Marchevski$\,$$\renewcommand{\thefootnote}{\fnsymbol{footnote}}\footnotemark[1]\renewcommand{\thefootnote}{\arabic{footnote}}^,\,$\footnotemark[13],
  K.~Massri,
 M.~Noy,
 V.~Palladino$\,$\footnotemark[27],
 M.~Perrin-Terrin$\,$\footnotemark[28]$^,\,$\footnotemark[29],
 J.~Pinzino$\,$\footnotemark[30],
 V.~Ryjov,
 S.~Schuchmann$\,$\footnotemark[13],
 S.~Venditti\\[2mm]

{\bf University of Birmingham, Birmingham, United Kingdom}\\
 T.~Bache,
 M. B.~Brunetti$\,$\footnotemark[31],
 V.~Duk$\,$\footnotemark[32],
 V.~Fascianelli$\,$\footnotemark[33],
 J. R.~Fry,
 F.~Gonnella,
 E.~Goudzovski,
 J.~Henshaw,
 L.~Iacobuzio,
 C.~Lazzeroni,
 N.~Lurkin$\,$\footnotemark[8],
 F.~Newson,
 C.~Parkinson$\,$\footnotemark[8],
 A.~Romano,
 A.~Sergi$\,$\footnotemark[34],
 A.~Sturgess,
 J.~Swallow\\[2mm]

{\bf University of Bristol, Bristol, United Kingdom}\\
 H.~Heath,
 R.~Page,
 S.~Trilov\\[2mm]

{\bf University of Glasgow, Glasgow, United Kingdom}\\
 B.~Angelucci,
 D.~Britton,
 C.~Graham,
 D.~Protopopescu\\[2mm]

{\bf University of Lancaster, Lancaster, United Kingdom}\\
 J.~Carmignani,
 J. B.~Dainton,
 R. W. L.~Jones,
 G.~Ruggiero$\,$$\renewcommand{\thefootnote}{\fnsymbol{footnote}}\footnotemark[1]\renewcommand{\thefootnote}{\arabic{footnote}}^,\,$\footnotemark[35]\\[2mm]

{\bf University of Liverpool, Liverpool, United Kingdom}\\
 L.~Fulton,
 D.~Hutchcroft,
 E.~Maurice$\,$\footnotemark[36],
 B.~Wrona\\[2mm]

{\bf George Mason University, Fairfax, Virginia, USA}\\
 A.~Conovaloff,
 P.~Cooper,
 D.~Coward$\,$\footnotemark[37],
 P.~Rubin\\[2mm]

\end{raggedright}
%
%
\setcounter{footnote}{0}
\renewcommand{\thefootnote}{\fnsymbol{footnote}}
\footnotetext[1]{Corresponding authors: F.~Brizioli, R.~Marchevski, G.Ruggiero. email: francesco.brizioli@cern.ch, \\
radoslav.marchevski@cern.ch, giuseppe.ruggiero@cern.ch}
\footnotetext[2]{Deceased}
\renewcommand{\thefootnote}{\arabic{footnote}}

\footnotetext[1]{Present address: CERN,  European Organization for Nuclear Research, CH-1211 Geneva 23, Switzerland}
\footnotetext[2]{Also at Laboratori Nazionali di Frascati, I-00044 Frascati, Italy}
\footnotetext[3]{Present address: Brookhaven National Laboratory, Upton, NY 11973, USA}
\footnotetext[4]{Also at Institute for Nuclear Research of the Russian Academy of Sciences, 117312 Moscow, Russia}
\footnotetext[5]{Present address: Faculty of Mathematics, Physics and Informatics, Comenius University, 842 48, Bratislava, Slovakia}
\footnotetext[6]{Also at TRIUMF, Vancouver, British Columbia, V6T 2A3, Canada}
\footnotetext[7]{Present address: Department of Astronomy and Theoretical Physics, Lund University, Lund, SE 223-62, Sweden}
\footnotetext[8]{Present address: Universit\'e Catholique de Louvain, B-1348 Louvain-La-Neuve, Belgium}
\footnotetext[9]{Present address: Institut f\"ur Kernphysik and Helmholtz Institute Mainz, Universit\"at Mainz, Mainz, D-55099, Germany}
\footnotetext[10]{Present address: Universit\"at W\"urzburg, D-97070 W\"urzburg, Germany}
\footnotetext[11]{Present address: European XFEL GmbH, D-22761 Hamburg, Germany}
\footnotetext[12]{Present address: University of Glasgow, Glasgow, G12 8QQ, UK}
\footnotetext[13]{Present address: Institut f\"ur Physik and PRISMA Cluster of Excellence, Universit\"at Mainz, D-55099 Mainz, Germany}
\footnotetext[14]{Also at Dipartimento di Fisica, Universit\`a di Modena e Reggio Emilia, I-41125 Modena, Italy}
\footnotetext[15]{Also at Faculty of Physics, University of Sofia, BG-1164 Sofia, Bulgaria}
\footnotetext[16]{Present address: University of Chinese Academy of Sciences, Beijing, 100049, China}
\footnotetext[17]{Present address: Institut am Fachbereich Informatik und Mathematik, Goethe Universit\"at, D-60323 Frankfurt am Main, Germany}
\footnotetext[18]{Also at Department of Industrial Engineering, University of Roma Tor Vergata, I-00173 Roma, Italy}
\footnotetext[19]{Also at Department of Electronic Engineering, University of Roma Tor Vergata, I-00173 Roma, Italy}
\footnotetext[20]{Also at Universit\`a degli Studi del Piemonte Orientale, I-13100 Vercelli, Italy}
\footnotetext[21]{Also at Universidad de Guanajuato, Guanajuato, Mexico}
\footnotetext[22]{Present address: Institute of Nuclear Research and Nuclear Energy of Bulgarian Academy of Science (INRNE-BAS), BG-1784 Sofia, Bulgaria}
\footnotetext[23]{Also at National Research Nuclear University (MEPhI), 115409 Moscow and Moscow Institute of Physics and Technology, 141701 Moscow region, Moscow, Russia}
\footnotetext[24]{Present address: DESY, D-15738 Zeuthen, Germany}
\footnotetext[25]{Present address: University of Lancaster, Lancaster, LA1 4YW, UK}
\footnotetext[26]{Present address: Charles University, 116 36 Prague 1, Czech Republic}
\footnotetext[27]{Present address: Physics Department, Imperial College London, London, SW7 2BW, UK}
\footnotetext[28]{Present address: Aix Marseille University, CNRS/IN2P3, CPPM, F-13288, Marseille, France}
\footnotetext[29]{Also at Universit\'e Catholique de Louvain, B-1348 Louvain-La-Neuve, Belgium}
\footnotetext[30]{Present address: INFN, Sezione di Pisa, I-56100 Pisa, Italy}
\footnotetext[31]{Present address: Department of Physics, University of Warwick, Coventry, CV4 7AL, UK}
\footnotetext[32]{Present address: INFN, Sezione di Perugia, I-06100 Perugia, Italy}
\footnotetext[33]{Present address: Center for theoretical neuroscience, Columbia University, New York, NY 10027, USA}
\footnotetext[34]{Present address: Dipartimento di Fisica dell'Universit\`a e INFN, Sezione di Genova, I-16146 Genova, Italy}
\footnotetext[35]{Also at  Dipartimento di Fisica e Astronomia dell'Universit\`a e INFN, Sezione di Firenze, I-50019 Sesto Fiorentino, Italy}
\footnotetext[36]{Present address: Laboratoire Leprince Ringuet, F-91120 Palaiseau, France}
\footnotetext[37]{Also at SLAC National Accelerator Laboratory, Stanford University, Menlo Park, CA 94025, USA}

\clearpage

\section{Introduction}
\label{sec:introduction}
The \pnnc decay is a Flavour Changing Neutral Current
(FCNC) process that proceeds at the lowest order in the Standard Model
(SM) through electroweak box and penguin diagrams, both dominated by
$t$-quark exchange. The quadratic Glashow-Iliopoulos-Maiani (GIM)
mechanism and the transition from a top to a down quark make this
process extremely rare. Using tree-level elements of the
Cabibbo-Kobayashi-Maskawa (CKM) matrix as external inputs, the SM
predicts the branching ratio to be BR
$ = (8.4\pm 1.0)\times 10^{-11}$~\cite{pnn_th}, where the uncertainty
is dominated by those of the CKM parameters $V_{cb}$ and $\gamma$. The
intrinsic theoretical uncertainty is 3.6$\%$, related to the
uncertainty of the NLO (NNLO) QCD corrections to the top (charm) quark
contribution~\cite{top_nlo_qcd,charm_nnlo_qcd} and NLO electroweak
corrections~\cite{top_nlo_ew}. The uncertainties due to the hadronic
matrix element governing the $K$--$\pi$ transition are negligible,
because they are evaluated using the precisely measured branching
ratio of \kethree, corrected for isospin-breaking and non-perturbative
effects calculated in~\cite{top_nlo_ew, lq_loops, lq_loops_1}.

The \pnnc decay is among the most promising modes to search for non-SM
signals in flavour physics, and it probes higher mass scales than
other rare meson decays.  The largest deviations from SM predictions
are expected in models with new sources of flavour violation, where
constraints from $B$ physics are weaker \cite{lh_tpar,
  warped_ed}. Models with tree-level FCNCs of left- or right-handed
chirality produce specific correlation patterns between the branching
ratios of \pnnz and \pnnc decay modes, which are constrained by the
value of the CP-violation parameter $\epsilon_K$
\cite{zprime,zprime_smeft}. Present experimental constraints limit the
range of variation within supersymmetric models
\cite{mssm,susy,mssm_tanbeta}. The \pnnc decay is also sensitive to
some aspects of lepton flavour non-universality \cite{lfuv} and can
constrain leptoquark models \cite{lq,lq1}.

The E787 and E949 experiments at the Brookhaven National Laboratory
studied the \pnnc decay using a kaon decay-at-rest technique and
measured BR=$(17.3^{+11.5}_{-10.5})\times 10^{-11}$
\cite{BNL09}. More recently, the NA62 experiment at the CERN Super
Proton Synchrotron (SPS) has measured the BR more precisely to be
$(4.8^{+7.2}_{-4.8})\times 10^{-11}$ using a decay-in-flight technique
and data recorded in 2016 \cite{na62_2016} and 2017 \cite{na62_2017}.

The NA62 experimental signature of the \pnnc decay is an incoming
$K^+$, an outgoing $\pi^+$, and missing energy and momentum. The
$K^+ \rightarrow \pi^+ X$ decay, where $X$ can be a scalar or a
pseudo-scalar particle, has the same signature. A search for a feebly
interacting particle $X$ gives access to physics Beyond the Standard
Model (BSM) at low energies. In a hidden-sector portal framework, the
$X$ particle mediates interactions between SM and hidden-sector fields
\cite{pbc_cern}. A scalar mediator $X$ can mix with the SM Higgs in
inflationary \cite{light_inflaton}, scale invariant
\cite{light_scalar}, and relaxion \cite{relaxion} models, which all
have cosmological implications. Models where $X$ is an axion, which
acquires mass from the explicit breaking of the Peccei-Quinn (PQ)
symmetry \cite{axion, axion1}, can be a signature of the PQ mechanism
and can solve the strong CP problem \cite{pq_strong_cp,
  pq_strong_cp_1}.  A QCD axion with a mass of
$\mathcal{O}(10^{-4})$~eV could be a dark matter candidate, and
specific axion models can also solve the SM flavour problem
\cite{axion_flavour}.  In a broader class of models, $X$ is considered
as an axion-like particle (ALP) that acts as a pseudoscalar mediator
\cite{pseudoS}.

The NA62 experiment collected its first data in 2016--2018. In the
following, the \pnnc analysis of the 2018 data set is described and
the combined result based on the full three-year sample is
presented. In addition, a search for the decay $K^+ \to \pi^+ X$ is
performed and upper limits are established in a particular scenario
\cite{pbc_cern}, where $X$ is a scalar particle.

\section{Beam line and detector}
\label{sec:beam-line-detector}
\begin{figure}[t]
  \begin{center}
 \includegraphics[width=1.0\textwidth]{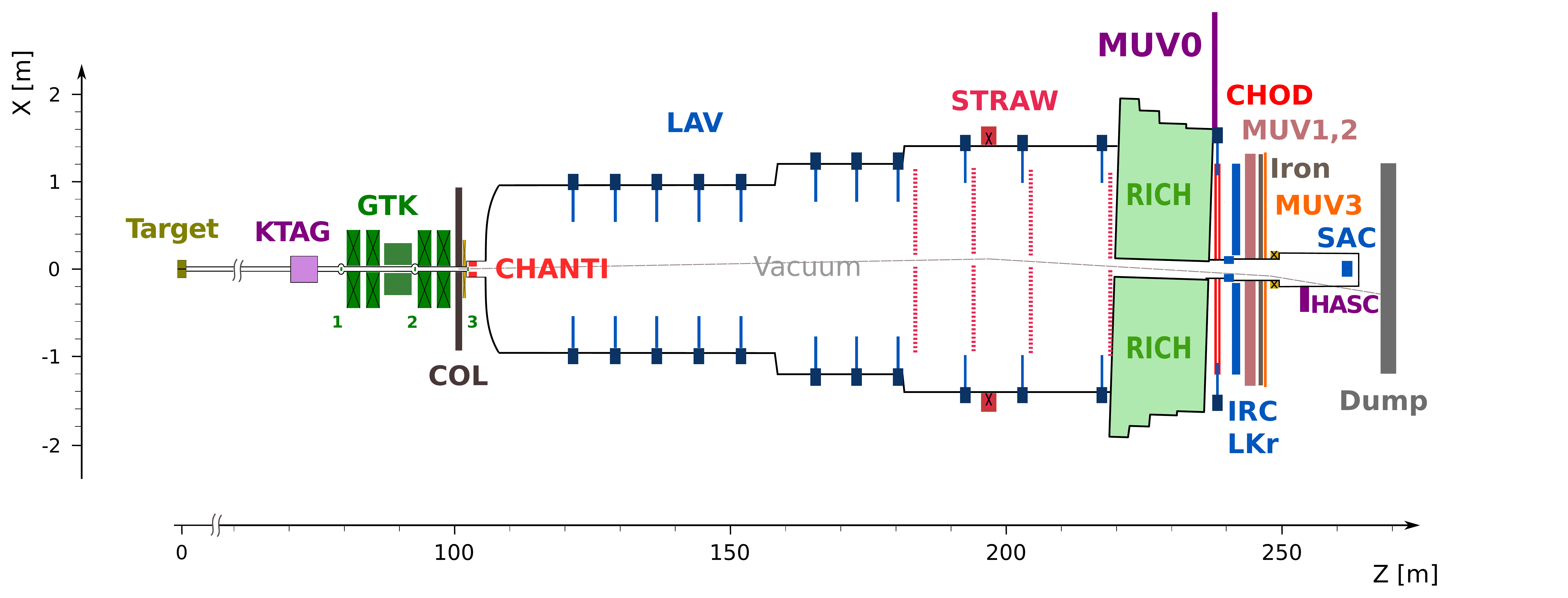}
 \caption{Schematic top view of the NA62 beam line and
   detector. Dipole magnets are displayed as boxes with superimposed
   crosses. The label ``COL'' denotes the collimator named ``final
   collimator'' in the text.  The label ``CHOD'' refers both to the
   CHOD and NA48-CHOD detectors.  Also shown is the trajectory of a
   beam particle in vacuum which crosses all the detector apertures,
   thus avoiding interactions with material. A dipole magnet between
   MUV3 and SAC deflects the beam particles out of the SAC
   acceptance.}
    \label{fig:layoutxz17}
  \end{center}
\end{figure}

The NA62 beam line and detector are sketched in
Figure~\ref{fig:layoutxz17} and a detailed description  can be
found in~\cite{na62det}.  The beam line defines the Z-axis of the
experiment's right-handed laboratory coordinate system.  The origin is
the kaon production target, and beam particles travel in the positive
Z-direction.  The Y-axis is vertical (positive up), and the X-axis is
horizontal (positive left).

The kaon production target is a 40~cm long beryllium rod.  A 400~GeV
proton beam extracted from the CERN SPS
impinges on the target in spills of three seconds effective duration.
Typical intensities during data taking range from $1.9$ to
$2.2\times10^{12}$ protons per pulse.
The resulting secondary
hadron beam of positively charged particles consists of 70\% $\pi^+$,
23\% protons, and 6\% $K^+$, with a nominal momentum of 75~GeV/$c$
(1\% rms momentum bite).

Beam particles are characterized by a differential Cherenkov counter
(KTAG) and a three-station silicon pixel matrix (Gigatracker, GTK,
with pixel size of $300 \times 300~\mu{\rm m}^2$).  The KTAG uses
N$_2$ gas at 1.75~bar pressure (contained in a 5~m long vessel) and is
read out by photomultiplier tubes grouped in eight sectors.  It tags
incoming kaons with 70~ps time-resolution.  The GTK stations are
located before, between, and after two pairs of dipole magnets (a beam
achromat), forming a spectrometer that measures beam particle
momentum, direction, and time with resolutions of 0.15~GeV/$c$,
16~$\mu$rad, and 100~ps, respectively.

The last GTK station (GTK3) is immediately preceded by a steel
collimator (final collimator, COL).  In the first part of the 2018
data taking period, as in 2016 and 2017, the inner variable aperture
of the 1~m thick collimator was typically set at
$66 \times 33~{\rm mm}^2$, and its outer dimensions were about
15~cm. It served as a partial shield against hadrons produced by
upstream $K^+$ decays. Shortly after the start of the 2018 data-taking
period, this collimator was replaced by a 1.2~m thick collimator with
outer dimensions $1.7 \times 1.8~{\rm m}^2$ and a central, race-track
shaped aperture of $76 \times 40~{\rm mm}^2$, designed to absorb all
hadrons emitted by upstream $K^+$ decays and passing outside its
aperture.

GTK3 marks the beginning of a 117~m long vacuum tank. The first 80~m
of the tank define a volume in which about 13\% of the kaons decay.  The
beam has a rectangular transverse profile of 52 $\times$ 24 mm$^2$ and
a divergence of 0.11~mrad (rms) in each plane at the decay volume
entrance.

The time, momentum, and direction of charged daughters of kaon
decays-in-flight are measured by a magnetic spectrometer (STRAW), a
ring-imaging Cherenkov counter (RICH), and two scintillator hodoscopes
(CHOD and NA48-CHOD).  The STRAW, consisting of four straw chambers,
two on each side of a dipole magnet, measures three-momenta with a
resolution, $\sigma_p / p$, between 0.3\% and 0.4\%.  The RICH, filled
with neon at atmospheric pressure, tags the decay particles with a
timing precision of better than 100~ps and provides particle
identification.  The CHOD, a matrix of tiles read out by silicon
photomultipliers, and the NA48-CHOD, comprising two orthogonal planes
of scintillating slabs reused from the NA48 experiment, are used for
triggering and timing, providing a time measurement with 200~ps
resolution.

Other sub-detectors suppress decays into photons or into multiple
charged particles (electrons, pions or muons) or provide complementary
particle identification.  Six stations of plastic scintillator bars
(CHANTI) detect, with 1~ns time resolution, extra activity, including
inelastic interactions in GTK3.  Twelve stations of ring-shaped
electromagnetic calorimeters (LAV1 to LAV12), made of lead-glass
blocks, are located inside and downstream of the vacuum tank to
achieve full acceptance for photons emitted by $K^+$ decays in the
decay volume at polar angles between 10 and 50~mrad.  A 27
radiation-length thick, quasi-homogeneous liquid krypton
electromagnetic calorimeter (LKr) detects photons from $K^+$ decays
emitted at angles between 1 and 10~mrad.  The LKr also complements the
RICH for particle identification.  Its energy resolution in NA62
conditions is $\sigma_E / E = 1.4\%$ for energy deposits of 25~GeV.
Its spatial and time resolutions are 1~mm and between 0.5 and 1~ns,
respectively, depending on the amount and type of energy released.
Two hadronic iron/scintillator-strip sampling calorimeters (MUV1,
MUV2) and an array of scintillator tiles located behind 80 cm of iron
(MUV3) supplement the pion/muon identification system.  MUV3 has a
time resolution of 400~ps.  A lead/scintillator shashlik calorimeter
(IRC) located in front of the LKr, covering an annular region between
65 and 135~mm from the Z-axis, and a similar detector (SAC) placed on
the Z-axis at the downstream end of the apparatus, ensure the
detection of photons down to zero degrees in the forward direction.
Additional counters (MUV0, HASC) installed at optimized locations
provide nearly hermetic coverage for charged particles produced in
multi-track kaon decays.

All detectors are read out with TDCs, except for LKr, MUV1 and MUV2,
which are read out with 14-bit FADCs.  The IRC and SAC are read out
with both.  All TDCs are mounted on custom-made (TEL62) boards, except
for GTK and STRAW, which each have specialized TDC boards.  TEL62
boards both read out data and provide trigger information.  A
dedicated processor interprets calorimeter signals for triggering.  A
dedicated board (L0TP) combines logical signals from the RICH, CHOD,
NA48-CHOD, LKr,  and MUV3 into a low-level trigger (L0) whose
decision is dispatched to sub-detectors for data
readout~\cite{na62trig}.  A software trigger (L1) exploits
reconstruction algorithms similar to those used offline with data from
KTAG, LAV, and STRAW to further reduce the data volume before storing
it on disk.

The data come from about $5\times10^5$ SPS spills accumulated during a
seven-month data-taking period in 2018, recorded at a mean
instantaneous beam particle rate of 500~MHz, measured event-by-event
using the number of signals recorded out-of-time in the GTK detector.
The average beam particle rate per spill was stable within $\pm$10\%
throughout the data-taking period, while the instantaneous value
showed fluctuations up to a factor of two around the average.

The data were collected using a trigger specifically setup for
the \pnnc measurement, called {\it PNN trigger}, concurrently with a
minimum-bias trigger.  The PNN trigger is defined as follows.  The L0
trigger requires a signal in the RICH to tag a charged particle.  The
time of this signal, called trigger time, is used as a reference to
define a coincidence within 6.3~ns of: a signal in one to four CHOD
tiles; no signals in opposite CHOD quadrants to suppress \pppc~decays;
no signals in MUV3 to reject \mn~decays; less than 30\,GeV deposited energy
  and no more than one cluster in the LKr to reject \pp~decays.
The L1 trigger requires: a kaon identified in KTAG; signals within 10
ns of the trigger time in at most two blocks of each LAV station; at
least one STRAW track corresponding to a particle with momentum below
50\,GeV$/c$ and forming a vertex with the nominal beam axis upstream
of the first STRAW chamber.  Events collected by the PNN trigger are
referred to as {\it PNN} events or {\it PNN} data.  The minimum-bias trigger is
based on NA48-CHOD information downscaled by a factor of 400.  In this case, the
trigger time is the time of the NA48-CHOD signal.  Data collected by
the minimum-bias trigger are used at analysis level to determine the
$K^+$ flux, to measure efficiencies, and to estimate backgrounds.
These data are called {\it minimum-bias} events or  {\it minimum-bias} data.

Acceptances and backgrounds are evaluated using a Monte Carlo (MC)
simulation based on the {\tt GEANT4} toolkit~\cite{geant4} to describe
detector geometry and response.  The $K^+$ decays are generated in the
kaon rest frame using the appropriate matrix elements and form
factors.  The simulation also includes a description of the
collimators and dipole and quadrupole magnets in the beam line,
necessary to accurately simulate the beam shape.  Certain aspects of
the simulation are tuned using input from data, namely signal
formation and readout detector inefficiencies.  Accidental activity is
added to the KTAG and GTK signals using the distribution of the
instantaneous beam particle rate measured with data, and a library of
GTK pileup hits built from GEANT4 simulations.  No accidental activity
is simulated in the detectors downstream of the last station of the
beam tracker.  Simulated data are subject to the same reconstruction
and calibration procedures as real data.

\section{Analysis method}
\label{sec:data-reconstr-calibr}

The experimental signature of the \pnnc decay consists of a $K^+$ with
4-momentum $P_{K}$ in the initial state and a $\pi^+$ with 4-momentum
$P_\pi$ and missing energy in the final state. The kinematic variable
used to discriminate between the signal and background $K^+$ decays is
the squared missing mass $m^{2}_{\rm miss} = (P_K - P_{\pi})^2$. This
variable is used to define the two regions used to search for the
signal (Region 1 and Region 2, as shown in Figure 2), and to separate
it from the other $K^+$ decay backgrounds.

The data set collected in 2018 is divided into two subsets, S1 and S2,
which correspond to the periods before (20$\%$ of the data set) and
after (80$\%$ of the data set) the installation of the new final
collimator COL.  The subset S2 is further divided into six categories
corresponding to equal 5~GeV$/c$ bins of pion momentum, $p_{\pi^+}$,
in the range 15--45~GeV$/c$. The subset S1 is considered as a separate
category and is integrated over $p_{\pi^+}$ due to its small size. A
dedicated selection is applied to each category, which improves signal
sensitivity. Data sets from 2016--2017, analyzed in~\cite{na62_2016}
and~\cite{na62_2017}, are added as two separate categories, each
integrated over $p_{\pi^+}$, for a total of nine categories.

The measurement of BR($K^+\rightarrow\pi^+\nu\bar{\nu}$) relies on the
calculation of the single event sensitivity ($SES$) and the background
evaluation for each category. The $SES$ is defined as
$1/(N_{K^+}\cdot \epsilon_{\pi\nu\bar{\nu}})$, where $N_{K^+}$ is the
effective number of $K^+$ decays in a pre-defined fiducial region and
$\epsilon_{\pi\nu\bar{\nu}}$ is the signal selection efficiency. The
\pp~decays selected from minimum-bias data are used as normalization
to compute $N_{K^+}$.  Signal and normalization decays share the same
selection defined by the presence of a single $\pi^+$ forming a vertex
with a parent $K^+$ inside the fiducial region. The rejection of extra
activity from photons or charged particles is applied only to the
signal selection. Control regions (Figure~\ref{fig:mm2_norm}) are used
to validate the background estimates. Control and signal regions are
masked until the completion of the analysis to avoid possible bias
during the optimization of the selection conditions. The \pnnc
branching ratio is obtained from a binned log-likelihood fit using the
signal acceptance and background expectation in each category.

\section{Event selection}
\label{sec:event-selection}

\begin{figure}[t]
  \begin{minipage}{0.7\textwidth}
    \includegraphics[width=1.0\textwidth]{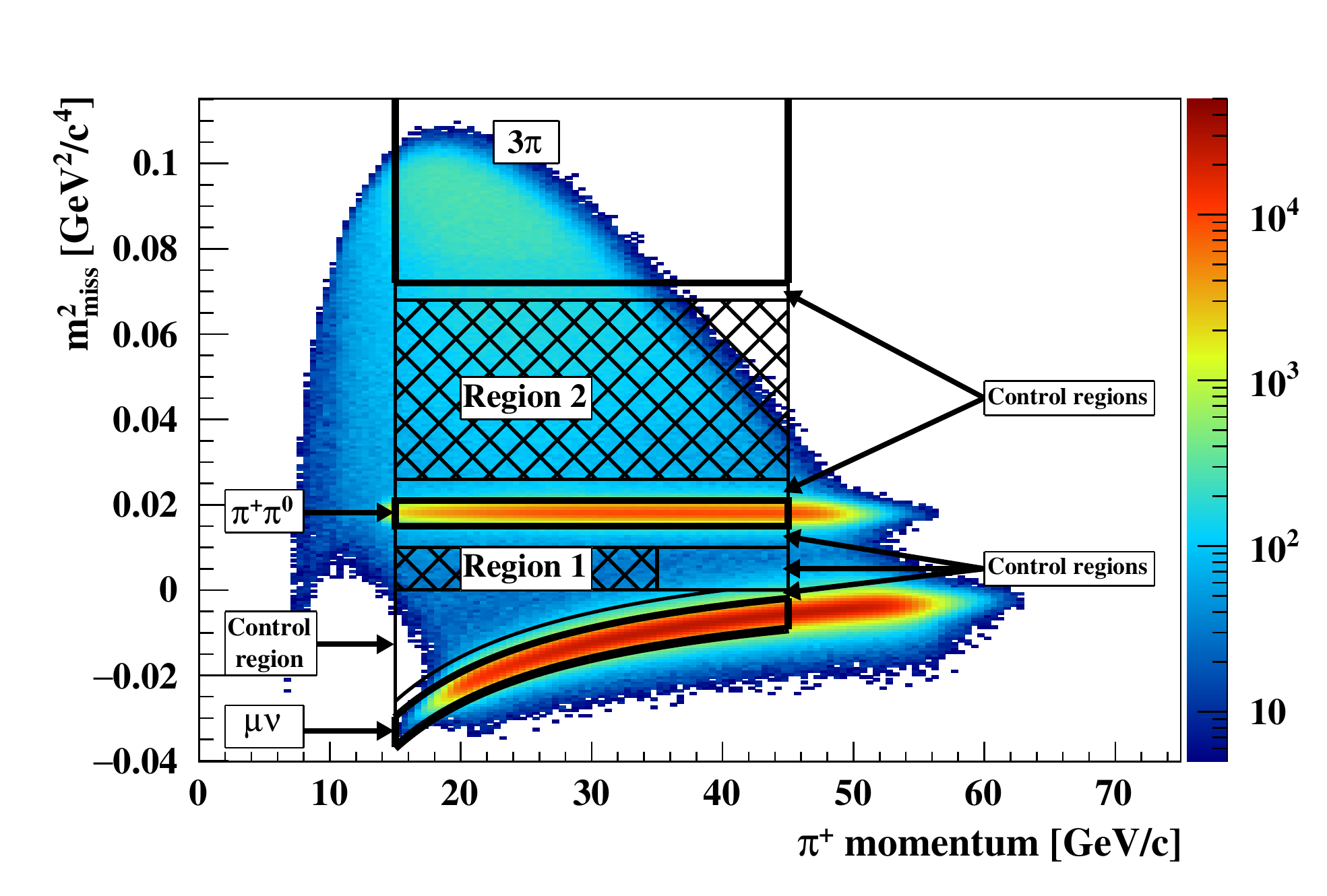}
  \end{minipage}
  \caption{Reconstructed $m^2_{\textrm{miss}}$ as a function of
    $p_{\pi^+}$ for minimum-bias events selected without applying
    $\pi^+$ identification and photon rejection, assuming the $K^+$
    and $\pi^+$ mass for the parent and decay particle, respectively.
    Signal regions 1 and 2 (hatched areas), as well as $3\pi$, $\pi^+\pi^0$,
    and $\mu\nu$ background regions (solid thick contours) are
    shown. The control regions are located between the signal and
    background regions.}
   \label{fig:mm2_norm}
\end{figure}

The signal and normalization channels both require the presence of a
downstream charged particle track identified as a $\pi^+$ and of a
parent $K^+$ track that forms a vertex in the fiducial volume. After
these common selection criteria, specific requirements define the
normalization and signal events.

\paragraph{Downstream charged particle:} One or two isolated STRAW
tracks are allowed in an event. If two STRAW tracks are present, the
one closest to the trigger time is selected. Events with a
negatively-charged STRAW track are rejected to remove \pppc~and
\kefour~decays.  The selected track must be within the RICH, CHOD,
NA48-CHOD, LKr, and MUV3 sensitive regions and must be spatially
associated to signals in the RICH, CHOD, NA48-CHOD, and LKr. The track
angle measured after the spectrometer magnet must be geometrically
compatible with the centre of the reconstructed RICH ring. Time
constraints are imposed on the associated signals in the RICH, CHOD
and LKr using the NA48-CHOD time as a reference. A STRAW track with
associated signals in CHOD, NA48-CHOD, RICH and LKr defines a
downstream charged particle.

\paragraph{Parent \boldmath$K^+$:} The parent $K^+$ of a selected
downstream charged particle is defined by the signal in KTAG with
time T$_{\rm KTAG}$ closest in time and within 2~ns of the downstream
particle, and a beam track in GTK with time T$_{\rm GTK}$ within 600~ps of
the KTAG signal and associated in space with the downstream track in
the STRAW. The association between the GTK, KTAG and STRAW signals
relies on a discriminant built from the time difference $\Delta$T =
T$_{\rm GTK}$ -- T$_{\rm KTAG}$ and the closest distance of approach
(CDA) of the downstream charged particle to the GTK track. The
templates for the $\Delta$T and CDA distributions of $K^+$ decays are
obtained from a dedicated sample of \pppc~decays, where the $K^+$ is
fully reconstructed using the pion momenta and directions measured by
the STRAW. The GTK track with the largest value of the discriminant
is then identified as the parent $K^+$. The same discriminant, but
with the RICH time of the downstream particle used as input instead of
T$_{\rm KTAG}$, strengthens the time coincidence and further rejects
accidental GTK tracks.  The selected $K^+$ must be consistent with the
nominal beam momentum and direction and the CDA must be less than
4~mm. No more than five reconstructed GTK tracks are allowed.  Events
with exactly one reconstructed GTK track are rejected if there are
additional hits within 300~ps in two GTK stations incompatible in
space with the selected $K^+$ track. The last condition reduces the
impact of GTK inefficiencies and of $K^+$ decays between the
second and third GTK stations.

\paragraph{Kaon decay:} The mid-point of the segment at the CDA of the
downstream charged particle to the parent $K^+$ defines the kaon decay
vertex. The Z position of the kaon decay vertex (Z$_{\rm vertex}$)
must be inside the region 110--165~m (110--170~m) for S1 (S2),
referred to as the fiducial volume (FV) in the following.  In
addition, the FV of the first $p_{\pi^+}$ bin (15--20~GeV$/c$) is
limited to 110--155~m to reduce the background from \pp~decays by
exploiting the correlation between momentum and Z$_{\rm vertex}$ of
the two-body decay kinematics. In the last two momentum bins
(35--45~GeV$/c$) the FV is limited to 110--160~m to suppress
\kefour~decays and upstream background, which dominate at high
momentum. Further Z$_{\rm vertex}$-dependent constraints are imposed
on the angle of the downstream charged particle. Its
backward-extrapolated position at the exit of the final collimator
(COL) must be outside a rectangular box with transverse dimensions
$100 \times 500~\textrm{mm}^2$. This condition is referred to as the
\textit{box cut} in the following and is applied to S1. The box cut is
needed to remove upstream $K^+$ decays entering the FV through the
aperture of the last achromat dipole, and leads to a 40\% signal
loss. For the sample collected with the new collimator, which
completely covers the aperture, the box cut size is reduced and most
of the acceptance loss is recovered.  This allows further optimization
of selection criteria using a Boosted Decision Tree (BDT) algorithm
for S2 instead of the cut-based approach applied to S1. The BDT inputs
are nine variables, which quantify the (X,Y,Z) coordinates and direction
of the downstream charged particle measured by the STRAW, and those
reconstructed at the decay vertex. The BDT is trained on an
out-of-time data sample enriched in upstream $K^+$ decays.  A cut on
the resulting BDT output value is chosen to provide the same
background rejection as a cut-based selection using the same
variables, while increasing the signal acceptance by 8$\%$.  Finally,
no signal must be present in the CHANTI detector to reduce the
contribution from $K^+$ inelastic interactions in GTK3.

\paragraph{Pion identification:} The $\pi^+$ identification uses
information from the calorimeters and the RICH and requires that no
signal is reconstructed in MUV3 within 7~ns of the $\pi^+$ time,
which reinforces the trigger condition.  A multivariate classifier
resulting from a BDT algorithm combines 13 variables describing the
energy associated with the $\pi^+$ in the calorimeters, the shape of the
clusters and the energy sharing between  LKr, MUV1 and MUV2.
Samples of $\pi^+,\,\mu^+$ and $e^+$ selected from 2017 data not
included in the present analysis are used for training.

The $\pi^+$ identification by the RICH uses two different approaches
to reconstruct a Cherenkov ring. In the first approach, the track
direction, as measured by the last two STRAW chambers, is used to
predict the position of the ring centre, and the expected ring radius
is calculated for each mass hypothesis ($e^+$, $\mu^+$, $\pi^+$,
$K^+$) using the track momentum.  For each mass hypothesis, a
likelihood is built by comparing the observed hit positions in the
RICH and a circle defined by the expected ring radius, and a cut is
applied on the value of the largest non-$\pi^+$ likelihood to remove
ambiguous events. In the second approach, the ring centre and radius
are determined by a $\chi^2$ fit to the hit positions, and the charged
particle mass is derived using the track momentum. A cut on the
measured mass is then applied to distinguish pions from
muons. Particle identification criteria with the calorimeters and RICH
are optimized separately for each data category to achieve the best
signal sensitivity.

\paragraph{Normalization selection:} The selection of the
normalization \pp~events is applied to minimum-bias data and requires
$0.010 < m^2_{\rm miss} < 0.026$~GeV$^2/c^4$ and
$15 < p_{\pi^+} < 45~$GeV$/c$. The width of the \textit{normalization
  region} is defined to be $\pm 8 \sigma$, where $\sigma$ is the
$m^2_{\rm miss}$ peak resolution. The resolution depends on the
$\pi^+$ track momentum and angular resolution, mostly due to multiple
scattering in the STRAW chambers. The MC simulation reproduces the
$m^2_{\rm miss}$ resolution to 10--20$\%$ and the discrepancy is
propagated as an uncertainty in the $SES$.

\paragraph{Signal selection:} The selection of the signal events is
applied only to PNN data and requires that no in-time photons or
additional charged particles are present. An in-time photon in the LKr
is defined as an energy cluster located at least 100~mm away from the
$\pi^+$ impact point and coincident in time with the $\pi^+$. The size
of the time coincidence window varies with the amount of deposited
energy and ranges from $\pm5$~ns below 1~GeV to $\pm50$~ns above
15~GeV. In-time photons in the LAV are identified if a signal is found
in any of the twelve LAV stations within 3~ns of the $\pi^+$ time. A
similar method is used for the IRC and SAC, where an in-time photon is
defined by either a TDC signal within 7~ns of the $\pi^+$ time or a
FADC signal of at least 1~GeV within 7~ns of the $\pi^+$ time.

Multi-charged particle rejection discriminates against interactions of
photons or charged particles in the RICH mirrors, and against \pppc~or
\kefour~decays with partially reconstructed STRAW tracks. The former
category is identified by the presence of isolated signals in at least
two of the CHOD, NA48-CHOD and LKr detectors as well as in-time
signals from the veto detectors MUV0 and HASC. The latter category is
characterized by the presence of tracks segments in the first two or
the last two STRAW chambers, which are consistent with a particle
coming from the kaon decay vertex.

The two \pnnc signal regions are defined in the $(p_{\pi^+},\,m^2_{\rm miss})$ plane as:
\begin{itemize}
\item[] \textbf{Region 1:} $0 < m^2_{\rm miss}< 0.010$~GeV$^2/c^4$ and
  $15~$GeV$/c< p_{\pi^+} < 35~$GeV/$c$;
\item[] \textbf{Region 2:} $0.026 < m^2_{\rm miss}< 0.068$~GeV$^2/c^4$ and
  $15~$GeV$/c< p_{\pi^+} < 45~$GeV/$c$.
\end{itemize}
Additional constraints are imposed on the $m^2_{\rm miss}$ value
using: the $\pi^+$ momentum extracted from the RICH ring measurement
in the $\pi^+$ mass hypothesis instead of the STRAW measurement; the
nominal beam momentum and direction instead of the $K^+$ track
measured by GTK. This reduces the kinematic tails due to multiple
scattering in the STRAW or wrong $K/\pi$ association.

The minimum momentum value is fixed at 15~GeV$/c$ by the RICH
threshold for efficient pion detection. The maximum value is fixed at
35~GeV$/c$ in Region 1, because the \mn~decay distribution approaches
the signal region at high momenta, and at 45~GeV$/c$ in Region 2 to
remove \kefour~and upstream backgrounds (see
Figure~\ref{fig:mm2_norm}).

\section{Single event sensitivity}
\label{sec:single-event-sens}
\begin{table}[t]
  \small
     \begin{tabular}[t]{l|c|c}
      \toprule
                                                           &  Subset S1             &  Subset S2 \\ \midrule
      $N_{\pi\pi} \times 10^{-7}$            & 3.14                       & 11.6 \\
      $A_{\pi\pi} \times 10^{2}$               & $7.62 \pm 0.77$  & $11.77 \pm 1.18$       \\
      $A_{\pi\nu\bar{\nu}} \times 10^{2}$ & $3.95 \pm 0.40$  & $6.37 \pm 0.64$\\
      $\epsilon^{\rm PNN}_{\rm trig}$      & $0.89 \pm0.05$  & $0.89 \pm0.05$ \\
      $\epsilon_{\rm RV}$                       & $0.66 \pm 0.01$  & $0.66\pm0.01$\\
      \hline
      $SES \times 10^{10}$              & $0.54 \pm 0.04$ & $0.14 \pm 0.01$\\
      $N_{\pi\nu\bar{\nu}}^{\rm exp}$         & $1.56 \pm 0.10 \pm 0.19_{\rm ext}$ & $6.02 \pm 0.39 \pm 0.72_{\rm ext}$\\
      \bottomrule
     \end{tabular} \hfill
     \caption{ Inputs to the $SES$ evaluation, $SES$ values and
       numbers of expected SM signal events in the S1 and S2 subsets.}
     \label{tab:ses}
\end{table}

The following expression is used to compute the $SES$ value:
\begin{equation}
  \label{eq:ses}
 SES = \frac{\textrm{BR}(K^+ \rightarrow \pi^+\pi^0)  \cdot A_{\pi\pi}}{ D \cdot N_{\pi\pi} \cdot A_{\pi\nu\bar{\nu}}\cdot \epsilon_{\rm RV} \cdot  \epsilon_{\rm trig}^{\rm PNN}}.
\end{equation}
Here $N_{\pi\pi}$ is the number of selected
$K^+ \rightarrow \pi^+\pi^0$ normalization events;
BR$(K^+ \rightarrow \pi^+\pi^0)$ is the \pp~branching ratio
\cite{pdg}; $D=400$ is the downscaling factor of the minimum-bias
trigger; $A_{\pi\nu\bar{\nu}}$ and $A_{\pi\pi}$ are the signal and
normalization acceptances, respectively, evaluated with simulations;
$1 - \epsilon_{\rm RV}$ is the inefficiency resulting from the random
veto induced by the photon and multi-charged particle rejection due to
the presence of accidental activity in the detectors;
$\epsilon_{\rm trig}^{\rm PNN}$ is the efficiency of the PNN trigger
stream. The inputs to the $SES$ computation, the resulting $SES$
values, and the corresponding numbers of expected SM \pnnc events for
the S1 and S2 subsets, integrated over $p_{\pi^+}$, are summarized in
Table~\ref{tab:ses}.

The \pnnc decays are simulated using form factors derived from the
\kethree~ decay. The accuracy of the description of particle
identification and $K/\pi$ association dominates the uncertainties of
$A_{\pi\nu\bar{\nu}}$ and $A_{\pi\pi}$ in Table~\ref{tab:ses}, but
these effects cancel to first order in the ratio
$A_{\pi\pi}/A_{\pi\nu\bar{\nu}}$.  The relative contribution of
$\pi^0$ Dalitz decays, $\pi^0 \rightarrow e^+ e^- \gamma$, to the
$SES$ result is estimated to be 0.7$\%$ and is assigned as a
systematic uncertainty to $A_{\pi\pi}$. A systematic uncertainty of
3.5$\%$ is propagated to the $SES$ value to take into account the
quality of the description of $\pi^+$ interactions with the material
upstream of the LKr, as well as of the $m^2_{\textrm{miss}}$
distribution. The former effect is estimated using \pp~decays with two
photons selected in the LAV stations, and the latter by the study of
the data/MC agreement of simulated \pp~decays.

The PNN trigger efficiency is a product of L0 and L1 efficiencies. The
non-calorimetric L0 trigger efficiency is measured using a sample of
\pp~events selected from minimum-bias data by applying signal-like
selection criteria without tagging the $\pi^0$.  The additional
requirement that both photons are detected by the LAV stations defines
the sample used to measure the calorimetric L0 trigger efficiency.
The L1 trigger efficiency is measured using a \mn~sample selected from
minimum-bias data and a \pp~sample triggered by the PNN L0
condition. The 2$\%$ uncertainty of this efficiency is propagated to
the $SES$ measurement. To test the assumption that the L0 and L1
trigger efficiencies are uncorrelated, a sample of \mn~events is
obtained from minimum-bias data by applying the \pnnc selection
criteria except for RICH $\pi^+$ identification.  The same selection
is applied to PNN data and the number of events in the $\mu\nu$ region
is compared to the number of minimum-bias events after correction for
the measured trigger efficiency and the downscaling factor $D$. The
observed discrepancy of up to 5$\%$, stable across the whole 2018
period, is propagated to the $SES$ measurement.

The random-veto parameter $\epsilon_{\rm RV}$ is measured using a sample of
\mn~decays from minimum-bias data, selected similarly to \pnnc decays
but with inverted particle identification criteria ($\mu^+$ instead of
$\pi^+$) and in the $\mu\nu$ $m^2_{\rm miss}$ region. The fraction of
events left after applying the photon and multi-charged particle
rejection is measured to be $\epsilon_{\rm RV} = 0.66 \pm 0.01$,
including a correction of $+0.02$ to account for activity in LAV and
CHOD induced by the $\delta$-rays produced by muons in the RICH
mirrors, as calculated from simulation. The value of $\epsilon_{\rm RV}$
depends on the instantaneous beam intensity, and its uncertainty is
evaluated by extrapolating $\epsilon_{\rm RV}$ to zero intensity and
comparing with a MC simulation of \mn~decays.

\section{Background evaluation and validation}
\label{sec:background-studies}
Background contributions to the \pnnc final state can be identified
from two processes: $K^+$ decays inside the FV to a final state
different from the signal; \textit{upstream events} where a $\pi^+$
originates either from a $K^+$ decay or from an interaction between a
beam $K^+$ and the material upstream of the FV.

\begin{figure}[t]
   \begin{minipage}{0.48\textwidth}
    \includegraphics[width=1.0\textwidth]{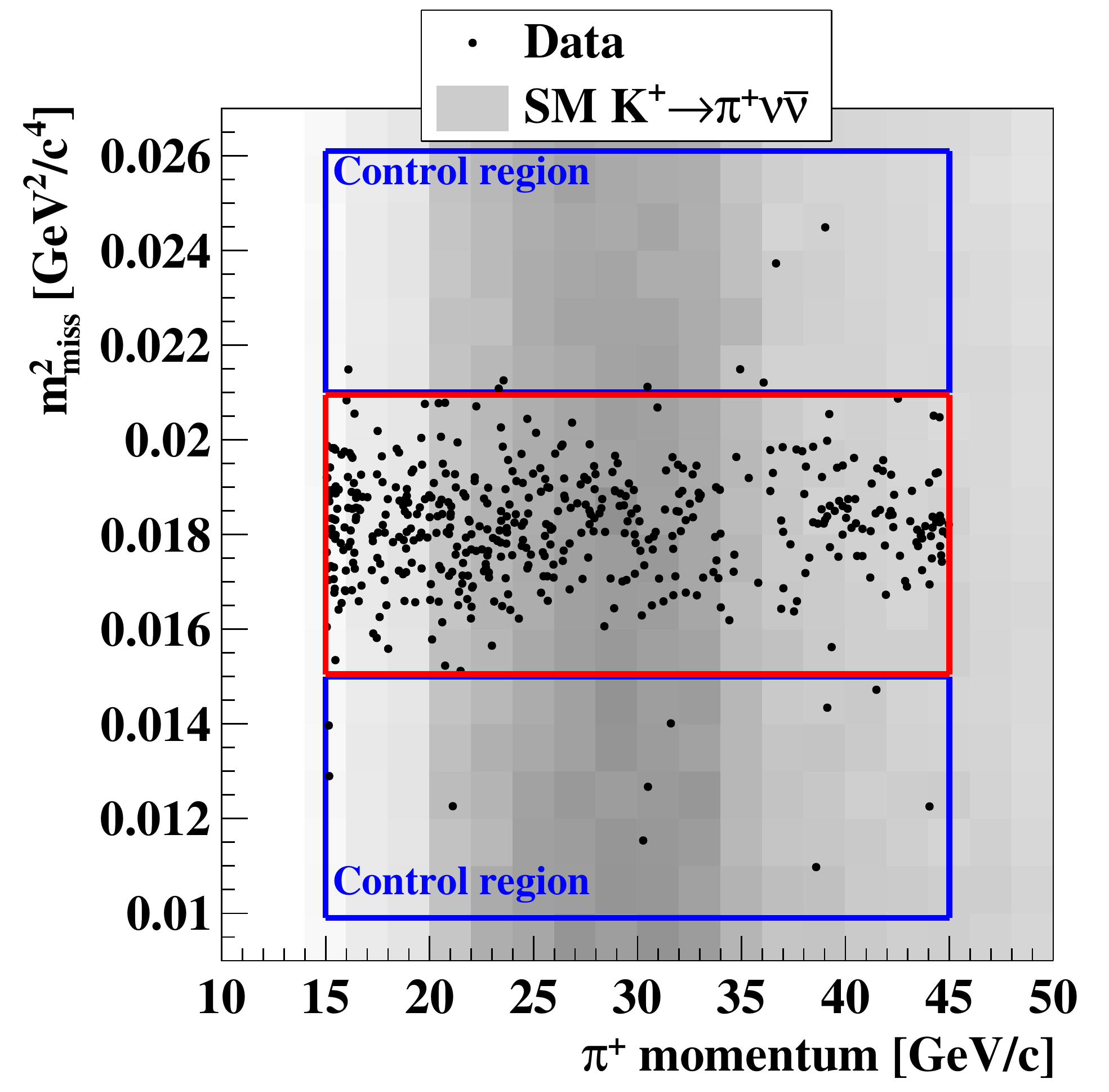}
  \end{minipage}\hfill
  \begin{minipage}{0.48\textwidth}\vspace{3mm}
    \includegraphics[width=1.03\textwidth]{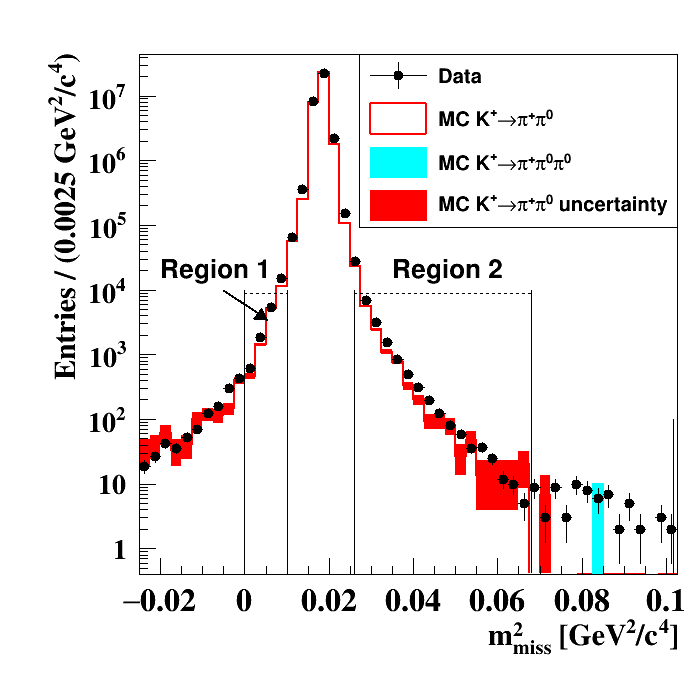}
  \end{minipage}\hfill
  \caption{Evaluation of the $\pi^+\pi^0$ background.  \textbf{Left:}
    Distribution in the ($p_\pi^+$, $m^2_{\textrm{miss}}$) plane of
    events in the $\pi^+\pi^0$ region and in the adjacent control
    regions after the complete signal selection is applied to the S1
    and S2 subsets. The intensity of the grey shaded area reflects the
    variation of the SM signal acceptance in the plane.
    \textbf{Right:} Data/MC comparison of the $m^2_{\textrm{miss}}$
    distribution of minimum-bias \pp~events selected by tagging the
    $\pi^0\rightarrow \gamma\gamma$ decay. This data sample is used to
    measure the \pp~kinematic factor $f_{\rm kin}$.}
  \label{fig:k2pi}
\end{figure}

The four main $K^+$ decay backgrounds are \pp, \mn, \pppc~and
\kefour. The first three enter the signal regions if
$m^2_{\textrm{miss}}$ is mis-reconstructed. The estimation of these
backgrounds relies on the assumption that $\pi^0$ rejection for \pp,
particle identification for \mn, and multi-charged particle rejection
for \pppc~are independent of the $m^2_{\rm miss}$ variable defining
the signal regions. After the \pnnc selection is applied, the expected
number of events in the signal or control regions is computed for each
category as:
\begin{equation}
  \label{eq:bgest}
  N^{\textrm{exp}}_{\textrm{decay}} = N_{\rm bkg}\cdot f_{\rm kin}(\rm region).
\end{equation}
Here $N_{\rm bkg}$ is the number of PNN-triggered events in the
$\pi^+\pi^0$, $\mu\nu$ or $3\pi$ background region and
$f_{\rm kin}(\rm region)$ is the fraction of events reconstructed in
the signal or control region for each decay mode. The values of the
kinematic factor $f_{\rm kin}(\rm region)$ are obtained: for \pp~ and
\mn~ by using minimum-bias data samples with dedicated selections; for
\pppc~by using simulated events.  Backgrounds from \kefour, \pgg and
semileptonic decays \klthree($l = e,\mu$) are evaluated only with
simulations.

\paragraph{\boldmath \pp:} After the \pnnc selection, 471 events are
observed in the $\pi^+\pi^0$ region (Figure~\ref{fig:k2pi}, left). The
kinematic factor is measured using a minimum-bias data sample with a
PNN-like selection applied.  The $\pi^0\rightarrow\gamma\gamma$ decay
is tagged by requiring exactly two photon clusters reconstructed in
the LKr, consistent with $m_{\pi^0}$ assuming that the photons
originate from the $K^+$ decay vertex, and no other activity in SAC,
IRC or LAV. This provides a background-free sample of \pp~decays
selected with no constraints on the $\pi^+$ kinematic variables. The
resulting $m^2_{\textrm{miss}}$ distribution is shown in
Figure~\ref{fig:k2pi}, right. The assumption that the $\pi^0$ tagging
is independent of the kinematics is tested by comparing the measured
kinematic factor in Region 1 with and without the $\pi^0$ tagging
applied. The disagreement of 3$\%$ is assigned as a systematic
uncertainty. The background estimates in the signal and control
regions are obtained using equation \ref{eq:bgest} and are validated
by comparing expected and observed numbers of events within the
control regions. The presence of radiative \ppgr decays from inner
bremsstrahlung increases the fraction of \pp~background. A correction
is applied to account for this effect using the simulated spectrum,
combined with a measurement of the energy-dependent single photon
detection efficiency of SAC, IRC, LAV and LKr on
data~\cite{NA62pi0inv}. The value of the correction represents 8$\%$
of the \pp~background and a 100$\%$ systematic uncertainty is assigned
to account for the precision of the simulation.

\paragraph{\boldmath \mn:} After the \pnnc selection, 14112 events are
observed in the $\mu\nu$ region. To take into account the correlations
between the RICH particle identification and the kinematic selection
criteria, $f_{\rm kin}(\rm region)$ is measured using a minimum-bias
data sample. The \mn~decays are selected applying signal-like
conditions, requiring the charged particle to satisfy the $\mu^+$
identification criteria in the calorimeters and the $\pi^+$
identification criteria in the RICH. Similarly to \pp, the kinematic
factor $f_{\rm kin}(\rm region)$ is measured by dividing the number of
events in the signal or control region by the number of events in the
$\mu\nu$ region.  The $\mu^+$ identification applied to the
minimum-bias sample suppresses the contribution from muons decaying in
flight as $\mu^+ \rightarrow e^+ \nu_e \bar\nu_{\mu}$ with $e^+$
misidentified as $\pi^+$. A simulation-driven correction of $+3\%$
(relative) is applied to the \mn~background estimation in Region 1, to
account for this effect. Events with real photon emission are included
in the measured $f_{\rm kin}$.

\paragraph{\boldmath \pppc:} The $m^2_{\textrm{miss}}$ distribution of
the three-body decay spans over a wide kinematic region. This
background is computed using an approach similar to the \pp~background
estimation while the kinematic factor $f_{\textrm{kin}}(\rm{region})$
is measured using MC simulations. A selection requiring only a match
between a $\pi^+$ in the final state and the parent $K^+$ is applied
to a sample of simulated \pppc~decays. To account for the resolution
of the $m^2_{\rm miss}$ variable, the factor
$f_{\textrm{kin}}(\rm{region})$ is computed in bins of
$m^2_{\rm miss}$. The $m^2_{\rm miss}$-dependent value of
$f_{\textrm{kin}}(\rm{region})$ is then multiplied by the number of
events in the corresponding $m^2_{\rm miss}$ bin of the $3\pi$ region
in data after the complete signal selection. The \pppc~background is
obtained after integrating the background estimates in each
$m^2_{\rm miss}$ bin.

\begin{figure}[t]
  \begin{minipage}{0.49\textwidth}
    \includegraphics[width=1.0\textwidth]{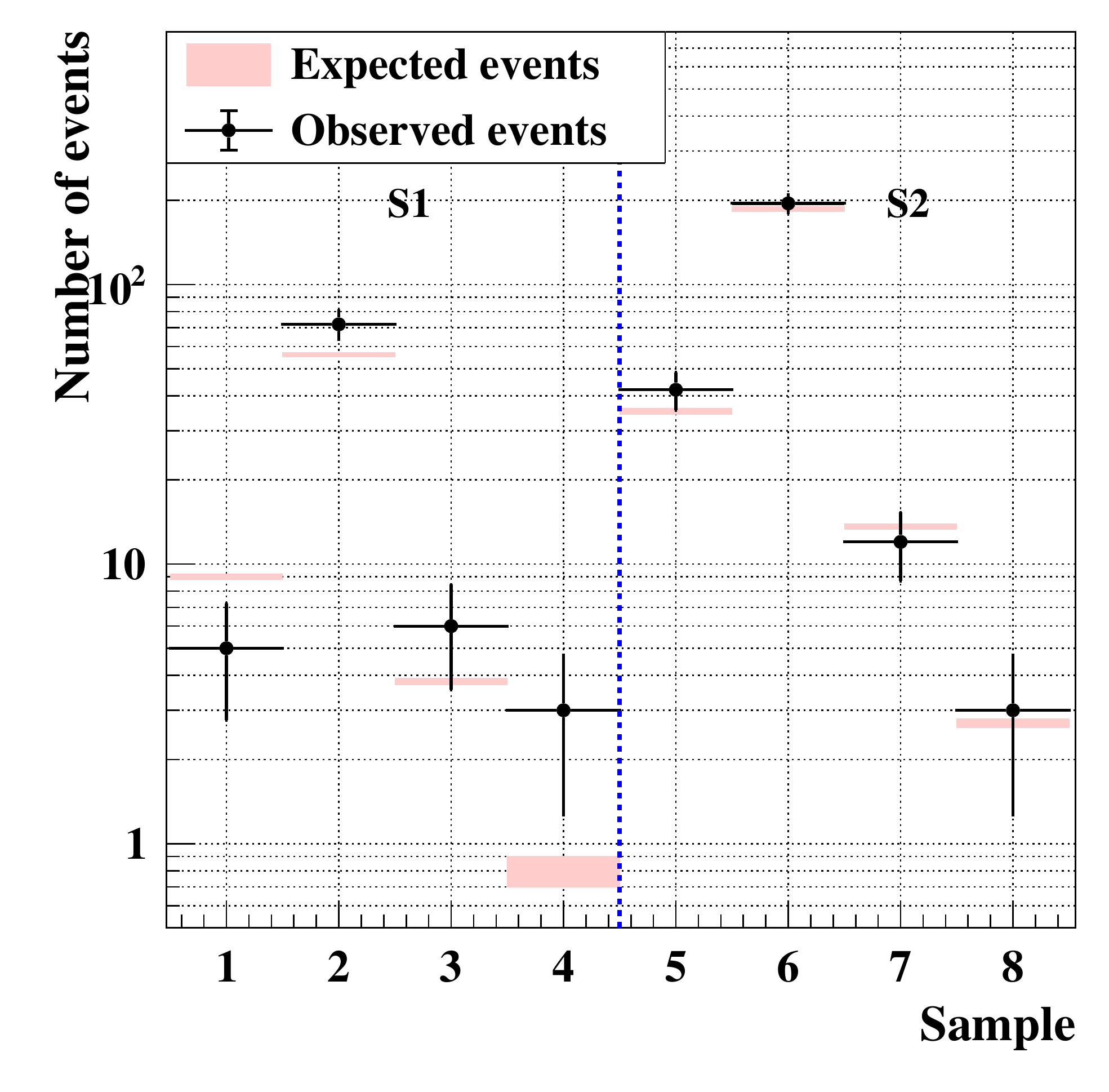}
  \end{minipage}\hfill
  \begin{minipage}{0.49\textwidth}
    \includegraphics[width=1.0\textwidth]{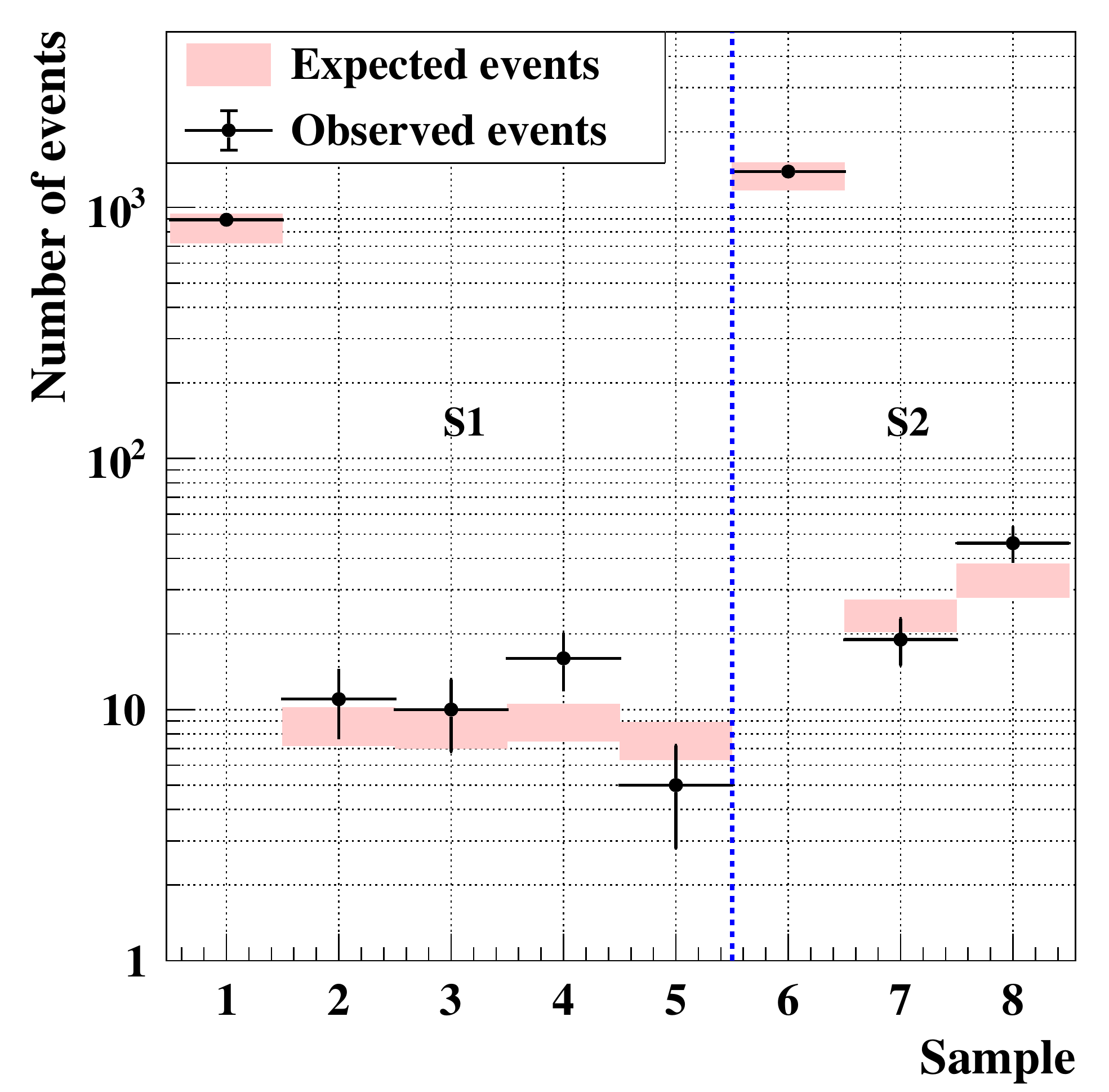}
  \end{minipage}\hfill
  \footnotesize
  \caption{ Validation of the \kefour~and upstream backgrounds.
  \textbf{Left:} Comparison between expected and observed number of events
  for the \kefour~validation samples in the S1 (1--4) and S2 (5--8) subsets.
    \textbf{Right:} Comparison between expected and observed number of events
    for the upstream background validation samples in the S1 (1--5) and S2 (6--8) subsets.}
  \label{fig:val}
\end{figure}

\paragraph{\boldmath \kefour:} This decay is characterized by large
values of $m^2_{\textrm{miss}}$ and contributes only to Region 2. The
contribution is suppressed by the $\mathcal{O}(10^{-5})$ branching
ratio, multi-charged particle rejection, particle identification and
kinematics. A sample of 2$\times 10^9$ MC simulated \kefour~decays
is used to estimate the background. The simulation is validated
by selecting four dedicated control samples. All samples are
statistically independent of the signal selection and are obtained by
inverting multi-charged particle rejection cuts or the charge sign of the
downstream charged particle. The $m^2_{\text{miss}}$ region used for
the validation is $0.030 < m^2_{\text{miss}} < 0.072$~GeV$^2/c^4$,
free from other background processes. Agreement is observed across all
samples as shown in Figure~\ref{fig:val}, left.

\paragraph{Other backgrounds from \boldmath $K^+$ decays:} The contributions
from \klthree$(l=\mu, e)$ and \pgg decays are found to be negligible
given the particle identification and photon rejection criteria
applied to the simulated samples. Upper limits of
$\mathcal{O}(10^{-3})$ and $\mathcal{O}(10^{-2})$ events are obtained
for the \klthree and \pgg contributions, respectively.

\paragraph{Upstream background:}

\begin{figure}[t]
  \begin{minipage}{0.48\textwidth}
    \includegraphics[width=1.0\textwidth]{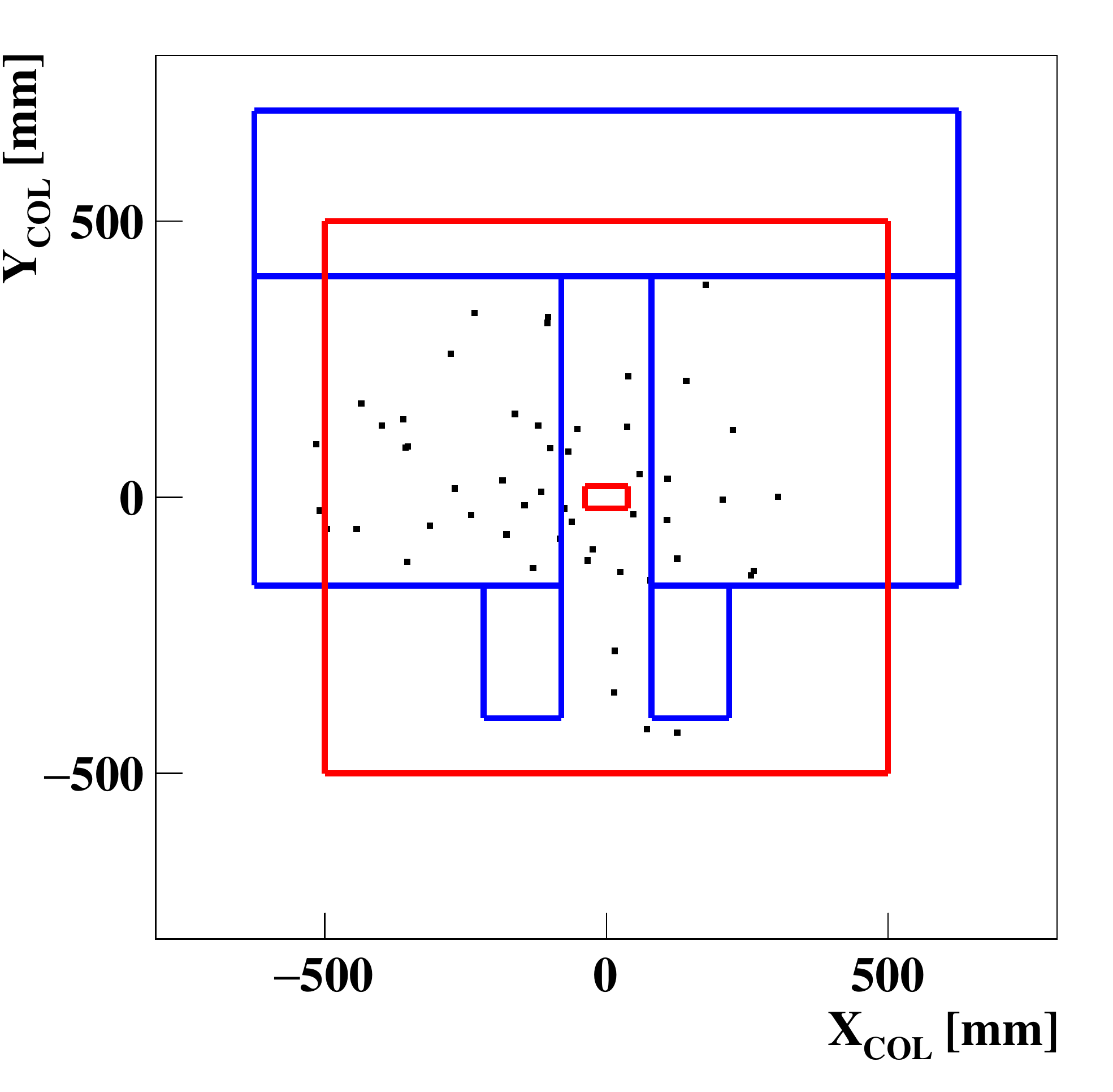}
  \end{minipage} \hfill
\begin{minipage}{0.48\textwidth}
    \includegraphics[width=1.0\textwidth]{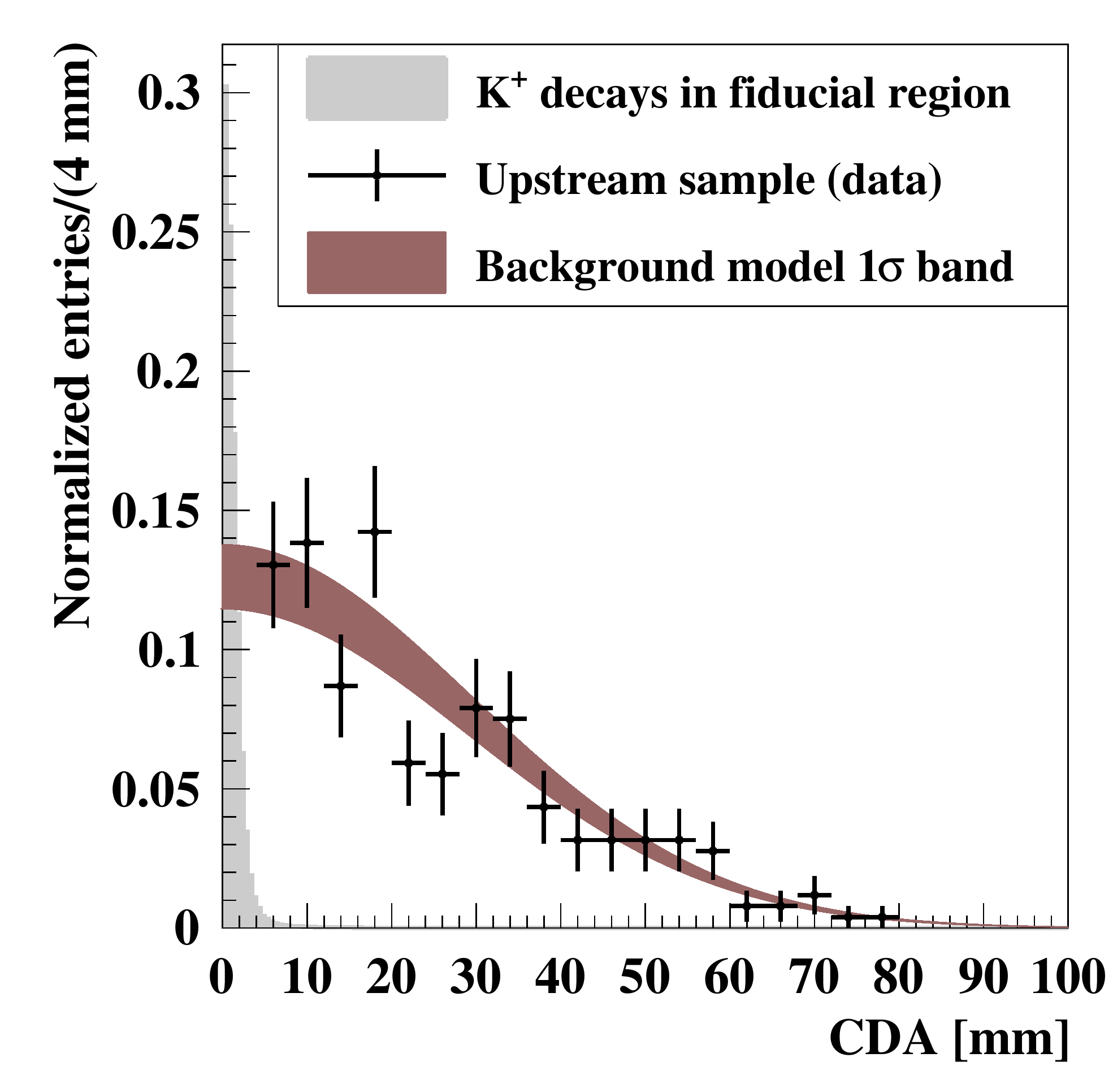}
  \end{minipage} \hfill
  \caption{Properties of upstream background events.  \textbf{Left:}
    Extrapolation of $\pi^+$ tracks of the upstream sample described
    in the text to the (X$_{\rm COL}$,\,Y$_{\rm COL})$ plane in the S2
    subset. The small (large) red rectangles correspond to the inner
    (outer) borders of the new collimator. The outline of the last
    dipole of the beam achromat is shown with blue solid lines.
    \textbf{Right:} CDA distribution of the events in the upstream
    sample shown on the left plot (black markers with error bars),
    compared to the CDA distribution extracted from data and its
    uncertainty (brown shaded area) and to the same distribution of
    $K^+$ decaying in the FV (grey shaded area).}
    \label{fig:up-bg}
\end{figure}
The background from upstream events receives contributions from two
types of processes: a $\pi^+$ from $K^+$ decays occurring between GTK
stations 2 and 3, matched to an accidental beam particle; a $\pi^+$
from interactions of a $K^+$ with the material in the beam line,
produced either promptly or as a decay product of a neutral kaon and
matched to the in-time $K^+$.  Studies on data and MC simulations
validate the above classification of upstream events.

The evaluation of the background from upstream events follows a
data-driven approach. A sample of PNN data is selected with all \pnnc
criteria applied, but requiring: CDA $>$ 4~mm; no $K/\pi$
association; $m^{2}_{\rm miss}$ value inside Regions 1 or 2. The
events selected define the \textit{upstream sample}: the distribution
of the $\pi^+$ tracks at the (X$_{\rm COL}$,\,Y$_{\rm COL})$ plane
(Section~\ref{sec:event-selection}) is shown in
Figure~\ref{fig:up-bg}, left. Contamination from $K^+$ decays in the
FV is at the per cent level and therefore negligible. The upstream
background is computed as the product of the number of events in the
upstream sample, $N_{\rm ups}$, and the probability, $P_{\rm mistag}$,
that an upstream event has \mbox{CDA $< 4$ mm} and satisfies the $K/\pi$
association criteria. The probability $P_{\rm mistag}$ depends on the
shapes of the distributions of CDA and of $\Delta$T.  The model of the
CDA distribution is extracted from the upstream sample by further
removing the veto condition on CHANTI signals and the criteria
suppressing pileup events in the GTK, which increase the number of
events in the upstream sample without biasing the CDA
(Figure~\ref{fig:up-bg}, right).  The probability $P_{\rm mistag}$ is
evaluated as a function of $\Delta$T by generating upstream-like
events in the (CDA,\,$\Delta$T) plane and applying the $K/\pi$
association with the CDA $<4$~mm condition. The expected background is
computed as

\begin{equation}
  \label{eq:ups-bg}
  N^{\rm exp}_{\rm upstream} = \sum_{i = 1}^{12} N_{\rm ups}(|\Delta{\rm T}_i|)\cdot P_{\rm mistag}(|\Delta{\rm T}_i|)\cdot f_{\rm scale}.
\end{equation}
The sum runs over the twelve 100~ps wide bins covering the
($-600,+600)$~ps region used to reconstruct the tracks in the GTK;
$ N_{\rm ups}(|\Delta{\rm T}_i|)$ is the number of events in the
upstream sample in the $\Delta$T bin $i$;
$P_{\rm mistag}(|\Delta{\rm T}_i|)$ is the mistagging probability;
$f_{\rm scale} = 1.15$ is a scaling factor that accounts for upstream
events with CDA $< 4$~mm not included in $N_{\rm ups}$.

In total, $N_{\rm ups}=9$ events are selected in S1 and
$N_{\rm ups}=38$ in S2, leading to an upstream background of
$N^{\rm exp}_{\rm upstream} = 3.3^{+0.98}_{-0.73}$ combining the S1
and S2 subsets. The uncertainty is dominated by the statistical
uncertainty of $N_{\rm ups}$. A systematic uncertainty of 20$\%$ is
added, related to the modelling of the CDA shape below 4~mm. A 15$\%$
systematic uncertainty is assigned to the value of $f_{\rm scale}$.

The background prediction is validated using a data sample enriched in
upstream events obtained by: removing the box cut (S1) or the cut on
the BDT output value (S2); selecting the $m^2_{\rm miss}$ value to be
either inside Regions 1 or 2, or in the region
$m^2_{\rm miss} < -0.05\textrm{~GeV}^{2}/c^{4}$ removing veto
conditions against pileup events in the GTK and interactions detected
by CHANTI. The selected events are distributed over five validation
samples in S1 depending on the $\pi^+$ position in the
(X$_{\rm COL}$,\,Y$_{\rm COL})$ plane, and on
$m^{2}_{\textrm{miss}}$. In the S2 subset, three validation samples
are obtained by either inverting the cut on the BDT output value or
selecting events in the unphysical region
$m^2_{\textrm{miss}} < -0.05$~GeV$^{2}/c^{4}$. The result of the
upstream background validation procedure is presented in
Figure~\ref{fig:val}, right.

\begin{figure}[t]
\begin{minipage}{0.59\textwidth}
    \includegraphics[width=1.0\textwidth]{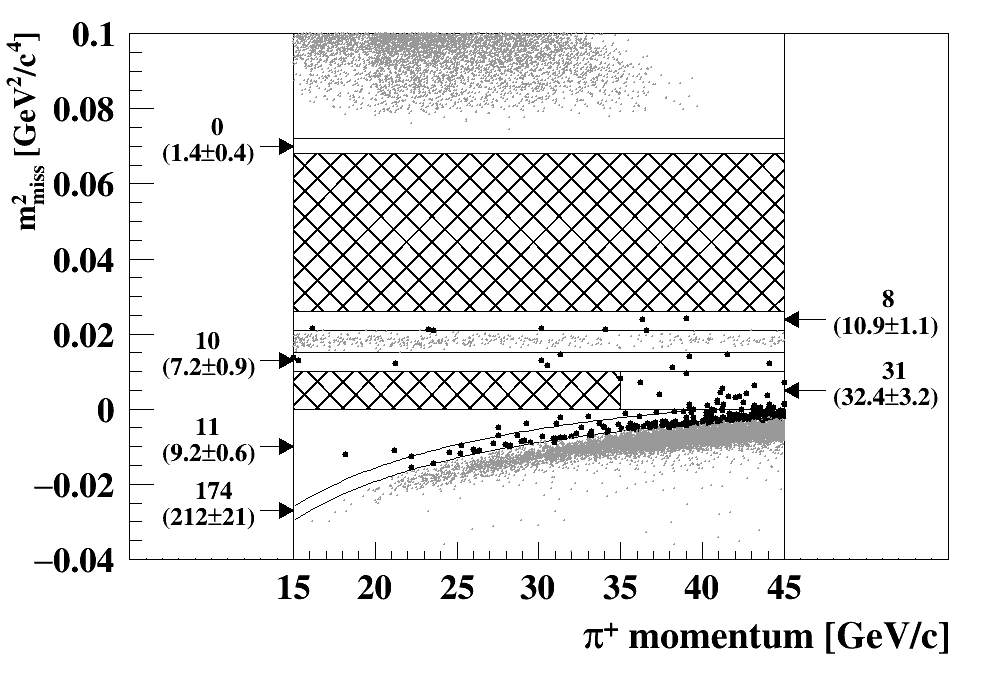}
  \end{minipage}
  \begin{minipage}{0.4\textwidth}
    \small
    \vspace{-0.8cm}
    \hspace{-0.2cm}
    \begin{tabular}{l|c|c}
      \toprule
      Background         & Subset S1            & Subset S2    \\\midrule
      $\pi^+\pi^0$         & $0.23\pm 0.02$  & $0.52\pm0.05$ \\
      $\mu^+\nu$           & $0.19\pm 0.06$  & $0.45\pm 0.06$\\
      $\pi^+\pi^-e^+\nu$   & $0.10\pm 0.03$  & $0.41\pm0.10$\\
      $\pi^+\pi^+\pi^-$    & $0.05\pm 0.02$  & $0.17\pm 0.08$\\
      $\pi^+\gamma\gamma$  & $<0.01$         & $<0.01$ \\
      $\pi^0l^+\nu$        & $<0.001$        &  $<0.001$\\
      Upstream     & $0.54^{+0.39}_{-0.21}$  & $2.76^{+0.90}_{-0.70}$\\ \midrule
      Total             & $1.11^{+0.40}_{-0.22}$& $4.31^{+0.91}_{-0.72}$\\
      \bottomrule
    \end{tabular}
  \end{minipage}

  \caption{Background predictions.  \textbf{Left:} Reconstructed
    $m^2_{\rm miss}$ as a function of $\pi^+$ momentum after applying
    the signal selection to the S1 and S2 subsets. Events in the
    background regions are displayed as light grey dots. The control
    regions, populated by the solid black markers, are adjacent to the
    background regions. The numbers next to these regions are the
    expected numbers of background events (in brackets) and the
    observed numbers (without brackets).  \textbf{Right:} Expected
    numbers of background events summed over Regions 1 and 2 in the
    2018 subsets. }
  \label{tab:bg}
\end{figure}

\paragraph{Background summary:}
The background prediction for the sum of all contributions described
above is validated in the six control regions located between the
signal and the $\pi^+\pi^0$, $\mu\nu$ and $3\pi$ regions.  After
unmasking the control regions the observed and expected numbers of
events are found to be statistically compatible across all control
regions (Figure~\ref{tab:bg}, left).  A summary of the background
estimates summed over Region 1 and Region 2 is presented in
Figure~\ref{tab:bg}, right for the two subsets S1 and S2 of the 2018
data.

\section{Results}
\label{sec:result}

\begin{figure}[h]
  \begin{minipage}{0.48\textwidth}
    \includegraphics[width=1.0\textwidth]{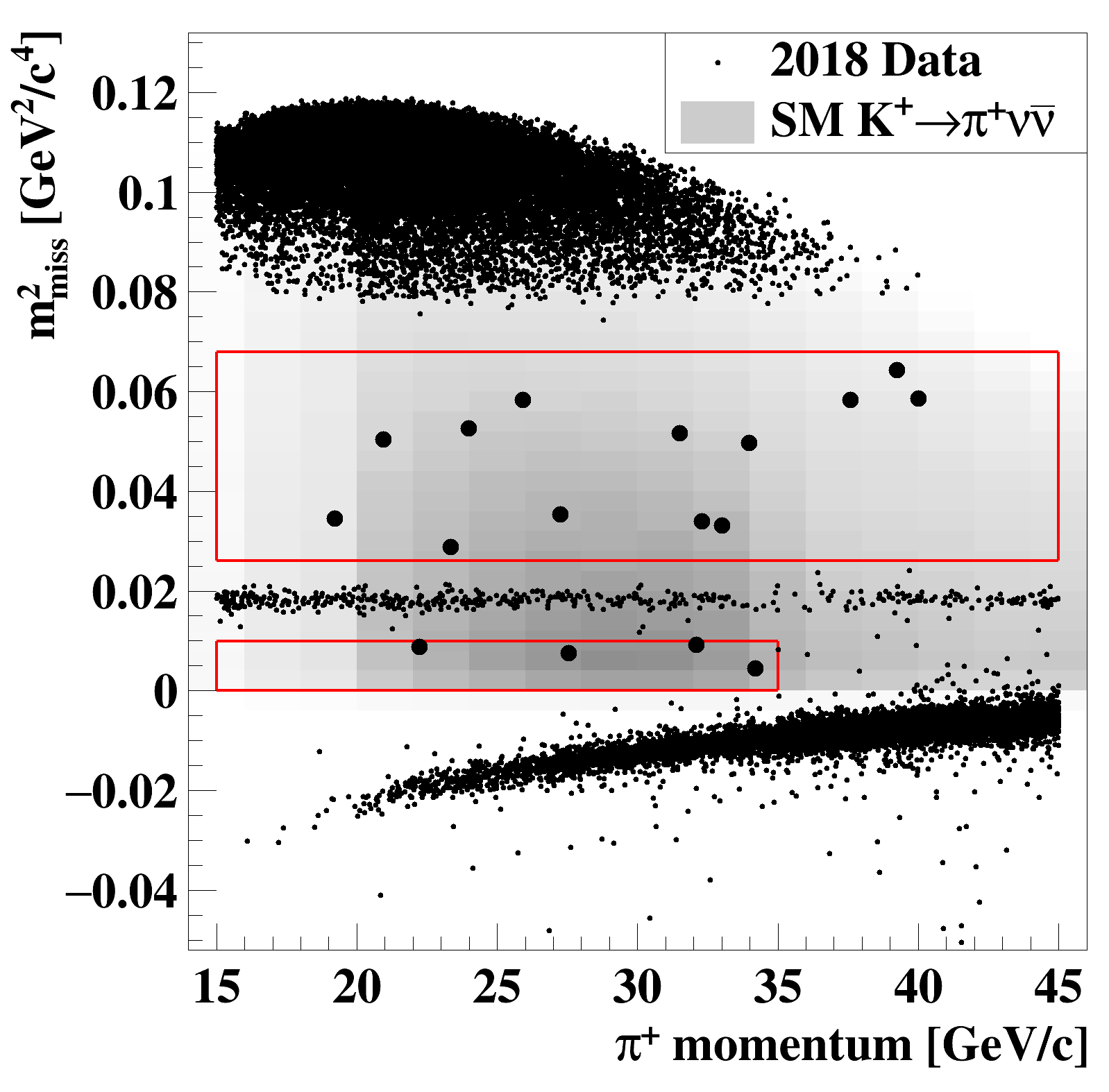}
  \end{minipage} \hfill
  \begin{minipage}{0.48\textwidth}
    \includegraphics[width=1.0\textwidth]{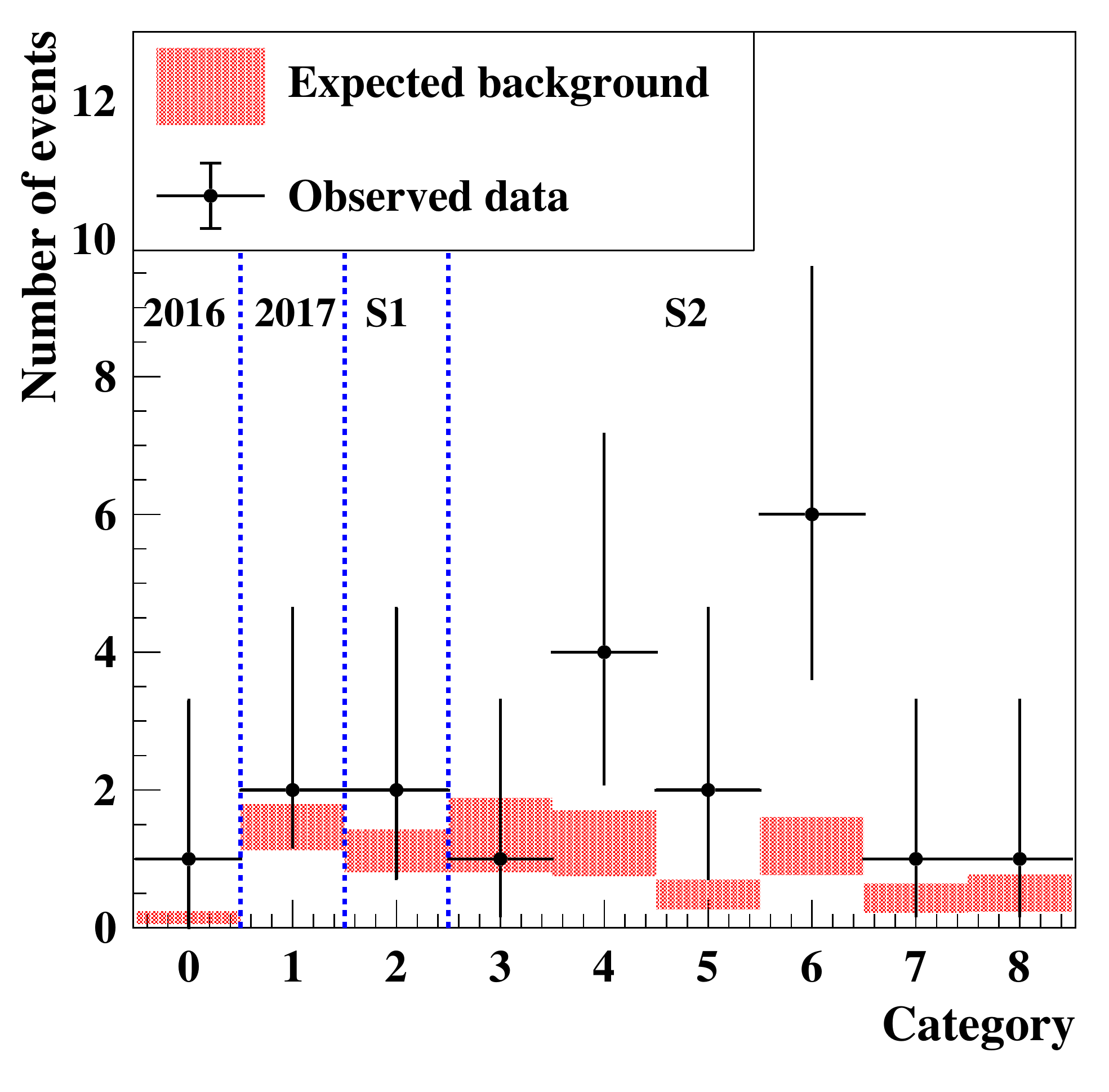}
  \end{minipage} \hfill
  \caption{The \pnnc candidate events in the 2018 (left) and
    2016--2018 (right) data sets.  \textbf{Left:} Reconstructed
    $m^2_{\rm miss}$ as a function of $\pi^+$ momentum for events
    satisfying the \pnnc selection criteria.  The intensity of the
    grey shaded area reflects the variation of the SM signal
    acceptance in the plane.  The two boxes represent the signal
    regions. The events observed in Regions 1 and 2 are shown together
    with the events found in the background and control
    regions. \textbf{Right:} Expected numbers of background events and
    numbers of observed events in the nine categories used in the
    maximum likelihood fit to extract the \pnnc branching
    ratio. Categories 0,1 and 2 correspond to 2016, 2017 and S1
    subsets, respectively. Categories 3 to 8 correspond to the six
    5~GeV$/c$ wide momentum bins of the S2 subset. The observed data
    for each category are indicated by black markers with Poissonian
    statistical errors. The shaded boxes show the numbers of expected
    background events and the corresponding uncertainties.}
    \label{fig:results}
\end{figure}

After unmasking the signal regions, four events are found in Region~1
and thirteen in Region~2, as shown in Figure~\ref{fig:results}, left.
In total, combining the results of the \pnnc analyses performed on the
2016, 2017 and 2018 data, 20 candidate events are observed in the
signal regions. The combined $SES$, and the expected numbers of signal
and background events in the 2016--2018 data set are:
\begin{align*}
  \label{eq:ses_npnn_run1}
  SES &= (0.839 \pm 0.053_{\rm syst})\times 10^{-11}, \\
  N^{\rm exp}_{\pi\nu\bar{\nu}} &= 10.01 \pm 0.42_{\rm syst} \pm 1.19_{\rm ext}, \\
  N^{\rm exp}_{\rm background} &= 7.03^{+1.05}_{-0.82}.
\end{align*}
The statistical uncertainties in the $SES$ and
$N^{\rm exp}_{\pi\nu\bar{\nu}}$ are negligible.  The above $SES$
corresponds to about $2.7 \times 10^{12}$ effective $K^+$ decays in
the fiducial volume.  The external error in
$N^{\rm exp}_{\pi\nu\bar{\nu}}$ stems from the uncertainty in the SM
prediction of BR$(K^+ \rightarrow \pi^+ \nu \bar{\nu})$. The
uncertainty in the background estimate is dominated by the statistical
uncertainty in the upstream background contribution.

A background-only hypothesis test is performed using as a test
statistic the likelihood ratio for independent Poisson-distributed
variables as prescribed in \cite{pdg}. A $p$-value of
$3.4\times 10^{-4}$ is obtained, corresponding to a signal
significance of 3.4 standard deviations.

The \pnnc branching ratio is determined using a binned maximum
log-likelihood fit to the observed numbers of events in the nine
categories comprising the NA62 data set (Figure~\ref{fig:results},
right). The parameter of interest is the signal branching ratio.  The
nuisance parameters are the total expected numbers of background
events in the signal regions and the single event sensitivities and
corresponding uncertainties, obtained separately for each of the nine
categories. For each category, the number of background events is
constrained to follow a Poisson distribution while the $SES$ follows a
Gaussian distribution with mean and sigma as estimated.  The resulting
branching ratio is
\begin{equation}
  \label{eq:3}
  \textrm{BR}(K^+\rightarrow\pi^+\nu\bar{\nu})= (10.6^{+4.0}_{-3.4}|_{\rm stat} \pm 0.9_{\rm syst})\times
  10^{-11}~\textrm{at}~68\%~\textrm{CL},
\end{equation}
compatible with the SM value within one standard deviation. The first
uncertainty is statistical, related to the Poissonian fluctuation of
the numbers of observed events and expected background, while the
second is systematic, resulting from the uncertainty in the signal and
background estimates.

This result is the most precise measurement of the \pnnc decay rate to
date and provides the strongest evidence so far for the existence of
this extremely rare process.

\section{Search for \boldmath $K^+  \to \pi^+  X$ decays}

\begin{figure}[t]
    \centering
    \begin{minipage}{0.49\textwidth}
        \includegraphics[width=1.0\textwidth]{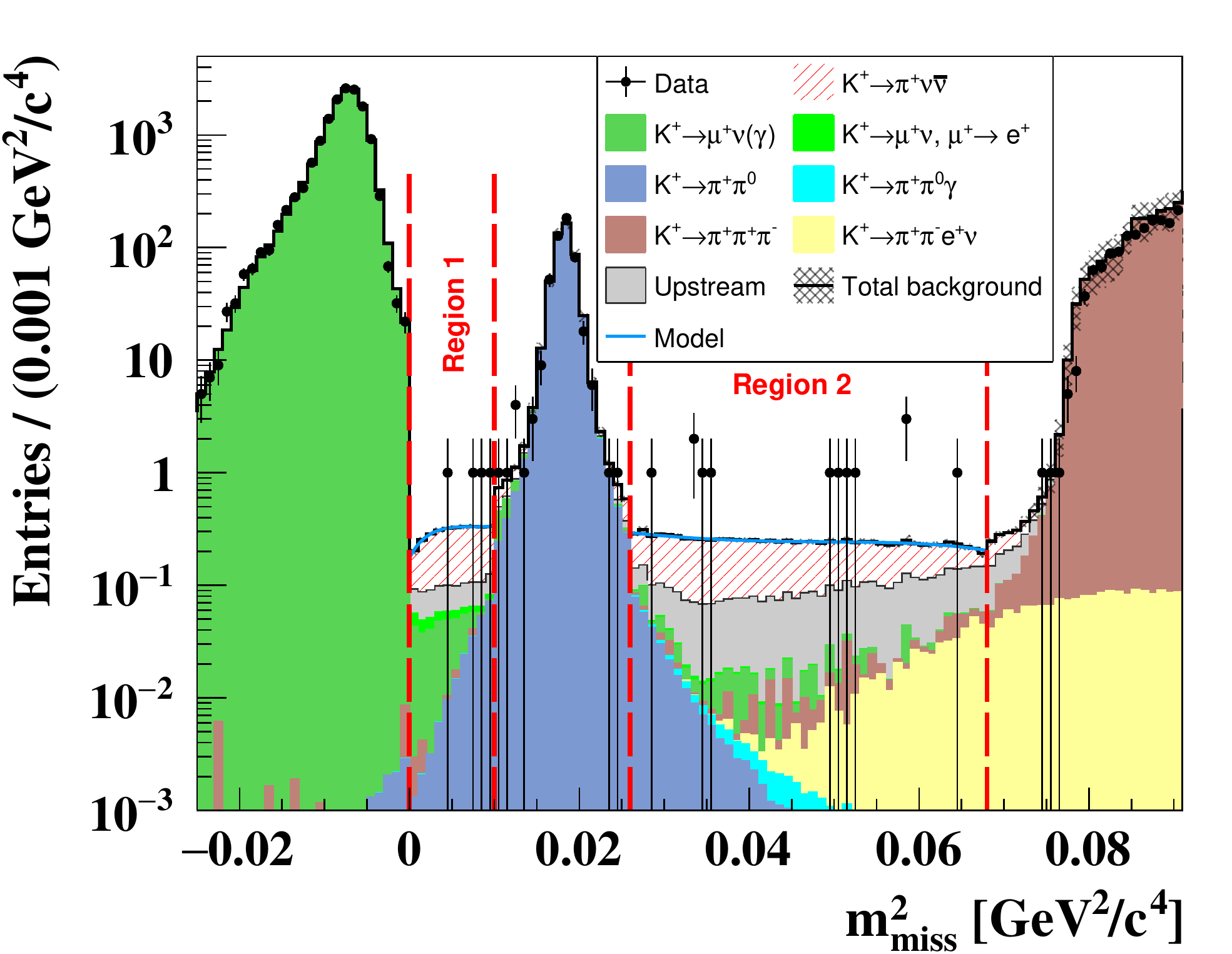}
    \end{minipage}
    \begin{minipage}{0.49\textwidth}
        \includegraphics[width=1.0\textwidth]{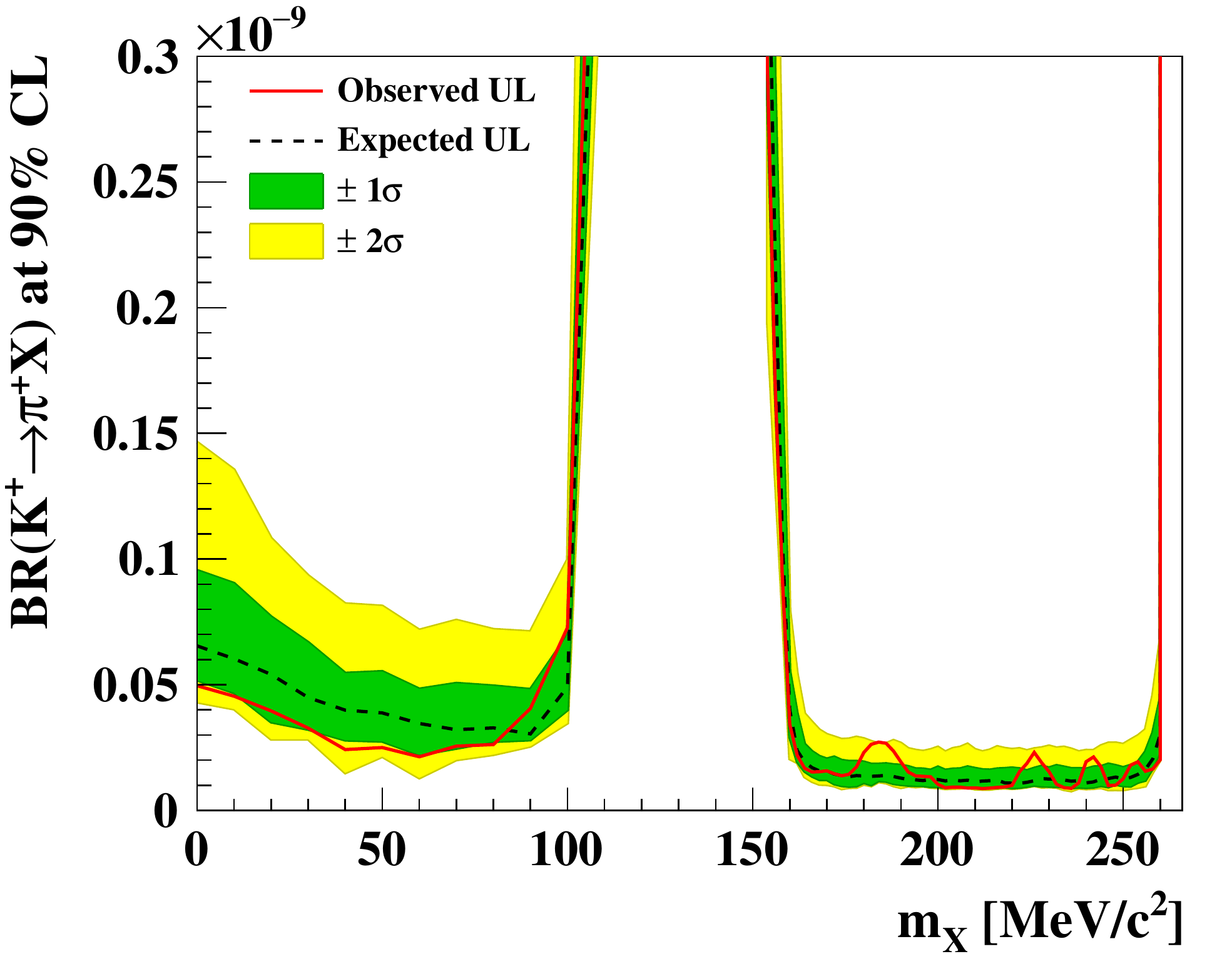}
    \end{minipage}
      \caption{ \textbf{Left:} Expected and observed number of
      events as a function of the reconstructed $m^2_{\rm miss}$ for the
      2018 data set.  \textbf{Right:} Upper limits on
      BR($K^+\rightarrow\pi^+ X$) for each tested $m_{\rm X}$ hypothesis, as obtained for the full 2016--2018 data set.
    }
    \label{fig:KpiX_ul}
  \end{figure}
  \begin{figure}[t]
    \begin{minipage}{0.49\textwidth}
      \includegraphics[width=1.0\textwidth]{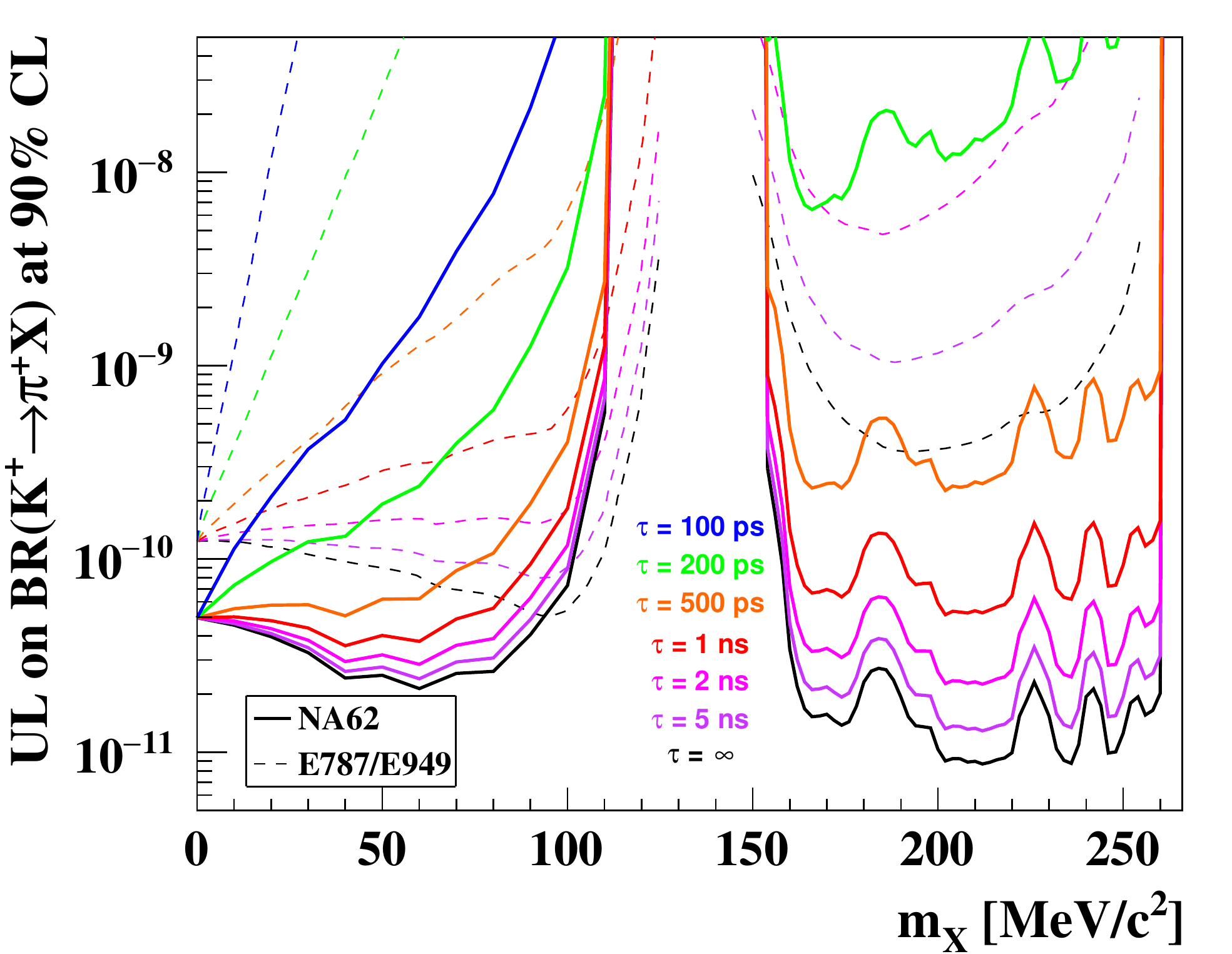}
    \end{minipage}
    \begin{minipage}{0.49\textwidth}
       \includegraphics[width=1.0\textwidth]{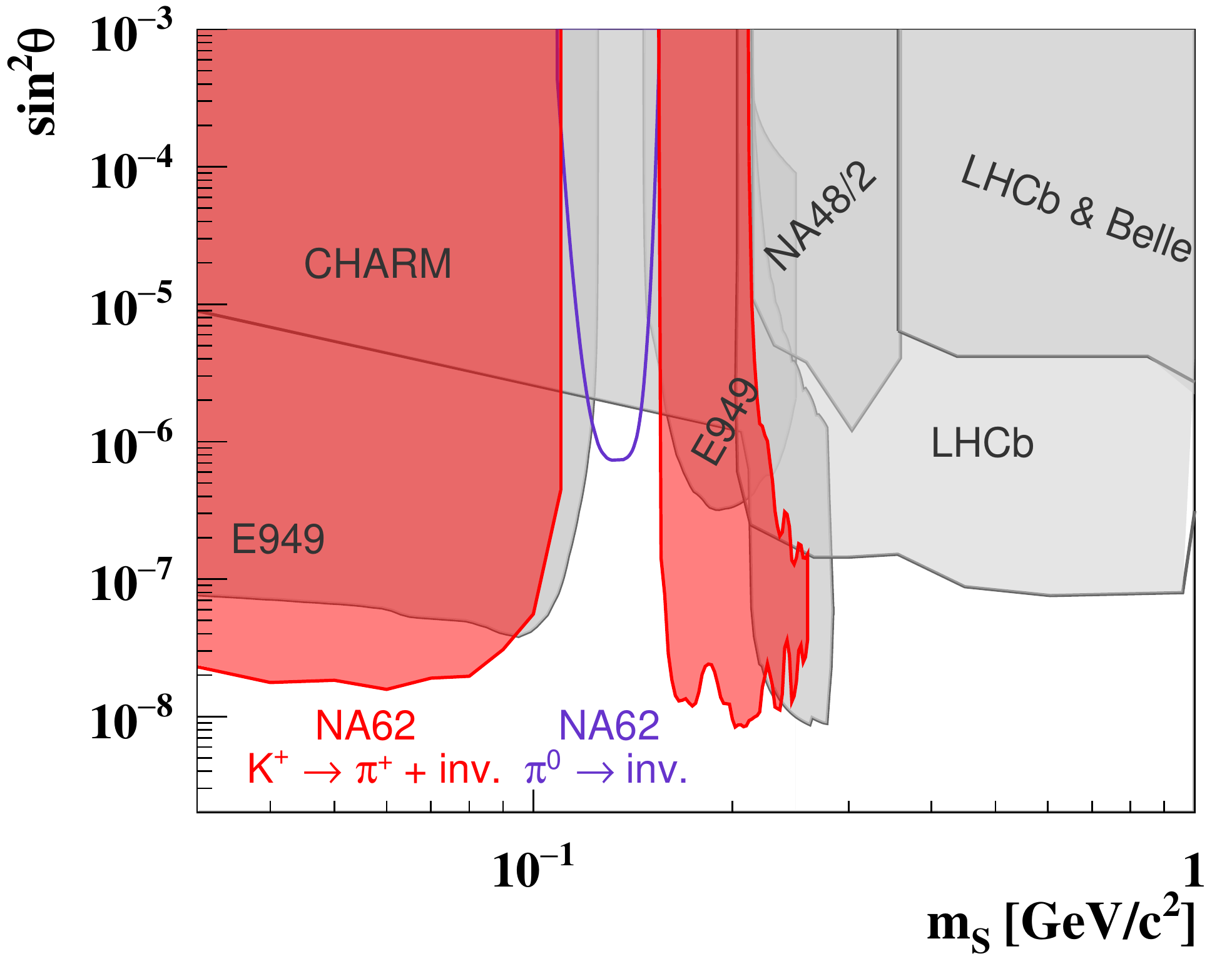}
       \end{minipage}
       \caption{ \textbf{Left:} Model-independent observed upper
         limits as functions of the $X$ mass and lifetime hypotheses.
         \textbf{Right:} Excluded regions of the parameter space
         $(m_{\rm S},\,\sin^{2}\theta)$ for a dark scalar, $S$, of the
         BC4 model~\cite{pbc_cern} decaying only to visible SM
         particles.  Exclusion bounds from the present search for the
         decay $K^{+}\rightarrow\pi^{+}X$ are labelled as
         ``$K^{+}\rightarrow\pi^{+}+\text{inv.}$'' and are shaded in
         red.  The constraints from the independent NA62 search for
         $\pi^{0}\rightarrow\text{\rm invisible}$
         decays~\cite{NA62pi0inv} are shown in purple.  Other bounds,
         shown in grey, are derived from the experiments
         E949~\cite{BNL09}, CHARM~\cite{Winkler},
         NA48/2~\cite{NA48Kpimumu},
         LHCb~\cite{LHCbBKmumu_a,LHCbBKmumu_b} and
         Belle~\cite{BelleBKmumu}.  }
    \label{fig:KpiX_interpretation}
\end{figure}

The existence of a new feebly interacting scalar or pseudo-scalar
particle, $X$, is foreseen in several BSM scenarios. If $X$ decays to
invisible particles or lives long enough to decay outside the
detector, the signature of a $K^{+}\rightarrow\pi^{+}X$ decay is the
same as that of the $K^{+}\rightarrow\pi^{+}\nu\bar{\nu}$ decay. The
two-body decay $K^{+}\rightarrow\pi^{+}X$ would result in a peak in
the reconstructed $m^2_{\rm miss}$ distribution, centred at the
squared value of the $X$ mass, $m_{\rm X}^{2}$.  Using the event
sample selected in the \pnnc measurement, a search for a peaking
signal in the 2016--2018 data set is performed following the procedure
detailed in~\cite{NA62KpiX17}.  The width of a signal peak is
determined by the resolution of the $m^2_{\rm miss}$ observable, which
decreases monotonically from $0.0012\text{~GeV}^{2}/c^{4}$ at
$m_{\rm X}=0$ to $0.0007\text{~GeV}^{2}/c^{4}$ at
$m_{\rm X}=260\text{~MeV}/c^{2}$.

The $SES$ is determined, for each $m_{\rm X}$, according to
equation~\ref{eq:ses}, by replacing $A_{\pi\nu\bar{\nu}}$ with the
acceptance for $K^{+}\rightarrow\pi^{+}X$ decays, which is obtained
from simulation.  Acceptance values for $X$ with finite lifetime,
$\tau_{\rm X}$, decaying to visible SM particles are estimated by
weighting simulated events by the probability that $X$ does not decay
upstream of MUV3.

The background contributions in searches for
$K^{+}\rightarrow\pi^{+}X$ decays are the same as for the
$K^{+}\rightarrow\pi^{+}\nu\bar{\nu}$ studies with the addition of the
$K^{+}\rightarrow\pi^{+}\nu\bar{\nu}$ decay itself, which becomes the
dominant background.  The expected distributions of background
processes as functions of $m^2_{\rm miss}$, evaluated as described in
Section~\ref{sec:background-studies} and assuming the SM description
for $K^{+}\rightarrow\pi^{+}\nu\bar{\nu}$ decay, are shown in
Figure~\ref{fig:KpiX_ul}, left. The largest uncertainty in the
estimated background comes from the SM \pnnc decay rate. The second
largest uncertainty comes from the modelling of the upstream
background distribution, which is statistically limited, and a
systematic uncertainty of up to $20\%$ is assigned to each
$m^2_{\rm miss}$ bin. The total background is modelled, as a function
of the reconstructed $m^2_{\rm miss}$, by polynomial functions fitted to
the expectations in Regions 1 and 2.

The search for $K^{+}\rightarrow\pi^{+}X$ decays is performed with a
fully frequentist hypothesis test using a shape analysis with
observable $m^2_{\rm miss}$ and an unbinned profile likelihood ratio
test statistic. Each $X$-mass hypothesis is treated independently
according to the CLs method~\cite{cls} to exclude the presence of a
signal with $90\%$~CL for the observed data. The statistical analysis
is performed using four categories corresponding to the 2016, 2017, S1
and S2 2018 subsets.

Under the assumption that the events observed in the signal regions
correspond to the known expected backgrounds, upper limits are
established on BR($K^{+}\rightarrow\pi^{+}X$) at $90\%$~CL for each
$X$-mass hypothesis. Results are displayed in
Figure~\ref{fig:KpiX_ul}, right for a stable or invisibly decaying
particle $X$. For $X$ decaying to visible SM particles, observed upper
limits are shown in Figure~\ref{fig:KpiX_interpretation}, left as a
function of $m_{\rm X}$ and for different values of $\tau_{\rm X}$.
These limits improve by a factor of four on those obtained from the
2017 data and improve over previous limits from the E949
experiment~\cite{BNL09} for most mass hypotheses.  The extension of
the FV for the S2 subset (Section~\ref{sec:event-selection}) with
respect to the analysis of the 2017 data set has improved
substantially the sensitivity to $X$ with shorter lifetimes.

An interpretation of these limits is presented for a scenario where
$X$ is a dark-sector scalar, which mixes with the Higgs boson
according to the mixing parameter
$\sin^{2}\theta$~\cite{pbc_cern,Winkler}. Constraints in the parameter
space of this scenario are shown in
Figure~\ref{fig:KpiX_interpretation}, right, under the assumption that
$X$ decays only to visible SM particles, with $\tau_{\rm X}$ inversely
proportional to the mixing parameter.

\section{Conclusions}
The NA62 experiment at CERN has analysed the data set collected in
2018, searching for the very rare $K^+ \rightarrow \pi^+ \nu\bar{\nu}$
decay, taking advantage of new shielding against decays upstream of
the kaon decay volume, and of improved reconstruction algorithms and
particle identification performance with respect to earlier data
sets. The statistical power was increased by three multiplicative
factors, one factor of 1.8 due to the larger number of effective kaon
decays, and two factors of 1.4, each due to better shielding and to
improved analysis technique. Combining the results obtained from the
whole 2016--2018 data set, a single event sensitivity of
$(0.839 \pm 0.053_{\rm syst})\times10^{-11}$ has been reached.  The
number of expected $K^+ \rightarrow \pi^+ \nu\bar{\nu}$ events in the
signal regions is $(10.01\pm0.42_{\rm syst}\pm1.19_{\rm ext})$,
assuming the Standard Model BR of ($8.4\pm 1.0) \times 10^{-11}$,
while $7.03 ^{+1.05}_{-0.82}$ background events are expected in the
same signal regions, mainly due to upstream background. After
unmasking the signal regions, twenty candidate events are observed,
consistent with expectation.  This leads to the branching ratio
BR$(K^+ \rightarrow \pi^+ \nu\bar{\nu}) = (10.6^{+4.0}_{-3.4} |_{\rm
  stat} \pm 0.9_{\rm syst}) \times 10 ^{-11}$~at~68\%~CL, which is the
most precise measurement to date.  In a background-only hypothesis, a
$p$-value of $3.4 \times 10^{-4} $ is obtained, which corresponds to a
3.4 standard-deviation evidence for this very rare decay.

This result is also interpreted in the framework of a search for a
feebly interacting scalar or pseudo-scalar particle $X$, produced in
the decay $K^+ \to \pi^+ X$ with the same experimental signature as
the dominant background process $K^+ \to \pi^+ \nu\bar{\nu}$. Upper
limits on the branching ratio at 90$\%$~CL of 3--6 $\times 10^{-11} $
are obtained for $m_{\rm X}$ masses in the range 0--110~MeV/$c^2$ and
$1 \times 10^{-11}$ for $m_{\rm X}$ masses in the range 160--260
MeV/$c^2$.  A particular model where $X$ is a dark-sector scalar
mixing with the Higgs boson has been explored, setting more stringent
constraints on the allowed region in the plane
($m_{\rm X},\,\sin^{2}\theta$), where $\theta$ is the mixing angle.

NA62 will continue taking data in 2021 with an upgraded detector
including beam line modifications, with the aim of further reducing
the upstream background, thus allowing for an improved signal
sensitivity.

\section*{Acknowledgements}
It is a pleasure to express our appreciation to the staff of the CERN laboratory and the technical
staff of the participating laboratories and universities for their efforts in the operation of the
experiment and data processing.

The cost of the experiment and its auxiliary systems was supported by the funding agencies of 
the Collaboration Institutes. We are particularly indebted to: 
F.R.S.-FNRS (Fonds de la Recherche Scientifique - FNRS), Belgium;
NSERC (Natural Sciences and Engineering Research Council), funding SAPPJ-2018-0017 Canada;
MEYS (Ministry of Education, Youth and Sports),  Czech Republic;
BMBF (Bundesministerium f\"{u}r Bildung und Forschung) contracts 05H12UM5, 05H15UMCNA and 05H18UMCNA, Germany;
INFN  (Istituto Nazionale di Fisica Nucleare),  Italy;
MIUR (Ministero dell'Istruzione, dell'Universit\`a e della Ricerca),  Italy;
CONACyT  (Consejo Nacional de Ciencia y Tecnolog\'{i}a),  Mexico;
IFA (Institute of Atomic Physics) Romanian CERN-RO No.1/16.03.2016 and Nucleus Programme PN 19 06 01 04,  Romania;
INR-RAS (Institute for Nuclear Research of the Russian Academy of Sciences), Moscow, Russia; 
JINR (Joint Institute for Nuclear Research), Dubna, Russia; 
NRC (National Research Center)  ``Kurchatov Institute'' and MESRF (Ministry of Education and Science of the Russian Federation), Russia; 
MESRS  (Ministry of Education, Science, Research and Sport), Slovakia; 
CERN (European Organization for Nuclear Research), Switzerland; 
STFC (Science and Technology Facilities Council), United Kingdom;
NSF (National Science Foundation) Award Numbers 1506088 and 1806430,  U.S.A.;
ERC (European Research Council)  ``UniversaLepto'' advanced grant 268062, ``KaonLepton'' starting grant 336581, Europe.

Individuals have received support from:
Charles University Research Center (UNCE/SCI/ 013), Czech Republic;
Ministry of Education, Universities and Research (MIUR  ``Futuro in ricerca 2012''  grant RBFR12JF2Z, Project GAP), Italy;
Russian Science Foundation (RSF 19-72-10096), Russia;
the Royal Society  (grants UF100308, UF0758946), United Kingdom;
STFC (Rutherford fellowships ST/J00412X/1, ST/M005798/1), United Kingdom;
ERC (grants 268062,  336581 and  starting grant 802836 ``AxScale'');
EU Horizon 2020 (Marie Sk\l{}odowska-Curie grants 701386, 754496, 842407, 893101).

\end{document}